\documentclass[10pt,a4paper]{article}
\usepackage{style}
\usepackage{amsmath,amssymb,epsfig,bm,amsfonts,yfonts,bbm,mathrsfs,hyperref,slashed}
\usepackage{pgfplots}
\usepackage[compat=1.1.0]{tikz-feynman}
\pgfplotsset{compat=newest}

\usepackage[scr=boondox,  
            cal=esstix]   
           {mathalpha}

\usepackage{pdfpages}
\usepackage{caption}
\usepackage{fontawesome5}
\usepackage{xcolor}

\definecolor{rot}{rgb}{0.7,0,0}
\definecolor{gruen}{rgb}{0,0.7,0}
\definecolor{blau}{rgb}{0,0,0.6}
\definecolor{hellblau}{rgb}{0.0, 0.7, 1.0}

\title{Second order and secondary functional renormalization group flow equations}

\author{Stefan Floerchinger}
\emailAdd{stefan.floerchinger@uni-jena.de}
\affiliation{Theoretisch-Physikalisches Institut, Max-Wien-Platz 1, 07743 Jena, Germany}

\abstract{The functional renormalization group provides a powerful non-perturbative framework for quantum field theory through the exact flow equation for the scale-dependent effective action. A mass-type regulator $R_k = k^2$ is particularly attractive because it preserves symmetries, locality, and unitarity (or Osterwalder--Schrader reflection positivity in the Euclidean domain). However, unlike momentum-dependent regulators, it does not tame all ultraviolet divergences: in four dimensions, flow equations for the effective potential exhibit logarithmic divergences. Motivated by Callan's perturbative renormalization scheme, where higher-order $\theta$-insertions improve ultraviolet convergence, we develop two classes of higher-order functional flow equations. First, we derive exact second-order flow equations by taking further derivatives of the standard flow equation with respect to $k$. These have either a one-loop structure involving the first derivative of the inverse propagator, or a mixed one-loop and two-loop structure. While exact and useful for incorporating two-loop effects and testing truncation schemes, they do not in general resolve the ultraviolet convergence problem. Second, we introduce a secondary flow method, where variations of an effective mass-squared parameter introduce an additional flow, with parameter $l$. The resulting secondary flow equations become ultraviolet finite once sufficiently many derivatives have been taken, and can be evaluated without an additional ultraviolet regulator. Integrating back to the standard flow requires integration constants that encode microscopic ultraviolet physics, analogous to relevant coupling constants.}

\begin{document}
\maketitle

\section{Introduction}

The functional renormalization group (FRG) provides a powerful framework for the non-perturbative study of quantum field theories. The central object is the exact, first-order flow equation for the scale-dependent effective action $\Gamma_k[\Phi]$, due to Wetterich \cite{Wetterich_1993_Exact} (see also \cite{Wetterich_1991_Average, Morris_1994_Exact, Ellwanger_1993_Flow, Reuter_1994_Effective}) which interpolates between the microscopic action at a high scale and the full quantum effective action in the infrared.

The functional renormalization group is closely related to other formulations of the exact renormalization group, most notably the Wilsonian flow equation of Polchinski \cite{Polchinski_1984_Renormalization} and its generalizations, going back to Wilson as well as Wegner and Houghton \cite{Wilson_1971_Renormalization, Wegner_1973_Renormalization, Rosten_2012_Fundamentals}. The different formulations are equivalent at the exact level, although they organize the degrees of freedom and the corresponding approximations in different ways. For comprehensive discussions of the functional renormalization group and its structural properties, see, e.g., \cite{Morris_1998_Elements, Aoki_2000_Introduction, Wetterich_2001_Effective, Polonyi_2003_Lectures, Pawlowski_2007_Aspects, Kopietz_2010_Introduction, Gies_2012_Introduction, Gurau_2014_Renormalization}.

The method employs a regulator function $R_k(p)$ which plays a double role: it suppresses quantum fluctuations in the infrared regime by acting like an additional contribution to squared mass for small momenta. But it also allows to effectively regularize the ultraviolet. This works, because the renormalization group flow is proportional to $\partial R_k(p)/\partial k$, and if that quantity decays fast enough for large momenta $p^2$ (in the Euclidean domain) one obtains ultraviolet finite flow equations.

The choice of regulator is not merely a technical matter. Although physical observables should be independent of the regulator in an exact treatment, approximate solutions generally retain a residual regulator dependence. This dependence provides useful information about the quality of truncations and motivates systematic regulator optimization procedures \cite{Litim_2000_Optimisation, Litim_2001_Optimized, Pawlowski_2007_Aspects, Reuter_2007_Functional, Delamotte_2012_Introduction, Schwenk_2012_Renormalization, Pawlowski_2017_Physics, Duclut_2017_Frequency, Knorr_2021_Exact, Braun_2021_Renormalization, DePolsi_2022_Regulator, Otto_2022_Regulator}.

One issue with the choice of a regulator function $R_k(p)$ is that it can break properties of a theory one would like to conserve. This is well studied for spacetime or internal symmetries, like rotations or global $O(N)$-symmetries, or also gauge symmetries. It is possible to avoid the breaking of many of these symmetries by constructing the regulator appropriately. For example, SO$(d)$,  the Euclidean version of Lorentz boost and rotation symmetry is preserved when $R_k(p)$ depends only on $p^2$.

The issue is more severe with unitarity, or Osterwalder-Schrader reflection positivity. Here it seems that there is no possibility to conserve it with any function $R_k(p)$ that is O$(d)$ invariant and decays for large $p^2$ \cite{Osterwalder_1973_Axioms, Osterwalder_1975_Axioms}, see also \cite{Floerchinger_2012_Analytic, Braun_2023_Renormalised, Ziebell_2023_Rigorous, Horak_2024_Scalar}. The reason is that multiplication with $R_k(p)$ in momentum space corresponds to a convolution in position space and the modified theories in the presence of $R_k(p)$ become then non-local. The only way out seems to be a constant, $R_k(p)= k^2$ (or a simple reparametrization thereof), for which locality and unitarity are preserved, at least for theories where a mass is anyway present).

Unlike other regulators, the mass-type regulator $R_k = k^2$ does not tame all ultraviolet divergences, however. In four dimensions, the flow equation for the effective potential and its derivatives exhibits logarithmic ultraviolet divergences \cite{Alexandre_2001_Functional, Braun_2023_Renormalised}. A central topic of this paper is to discuss these, and how they can be dealt with.

For this, we take some inspiration from Callan's perturbative renormalization scheme \cite{Callan_1970_Broken, Symanzik_1970_Small, Callan_1993_Introduction}, which is a close relative of functional renormalization with a mass-like regulator. In Callan's scheme, not only a first-order derivative with respect to the microscopic squared mass is used, but in situations where this is not enough to obtain finite expressions, also higher order derivatives, so-called $\theta$-insertions in his language, are being employed.

Motivated by this, the present work makes two major developments:
\begin{itemize}
	\item \textbf{Second-order flow equations:} We study higher order flow equations, which we obtain by taking a further derivative of the standard, first-order flow equation with one-loop structure with respect to the flow parameter $k$. This leads to new exact equations, which come in two variants: either as equations with a one-loop structure but where the right hand side depends itself on the first derivative of the full inverse propagator with respect to the flow parameter, or as flow equations with terms of one-loop and two-loop character. These equations are highly interesting by themselves, but they do not in general solve the ultraviolet convergence problem.
	\item \textbf{Secondary flow equations:} Here the idea is to take further derivatives of the standard, first-order flow equation with respect to the effective squared mass, or an additional regulator term inserted by hand. This leads to a secondary flow, or flow of the flow, with an additional parameter $l$. The secondary flow equations obtained in this way are indeed ultraviolet finite when enough derivatives with respect to $l$ have been taken. In the integration of the secondary flow one encounters integration constants, however, which are new parameters of the theory, that depend on microscopic physics, very much as relevant coupling constants.
\end{itemize}

\paragraph{This paper is organized as follows.} In section~\ref{sec:FirstOrderFunctionalFlowEquation} we review the standard first-order functional flow equations for the Schwinger functional, the effective action, and higher-order correlation functions. Section~\ref{sec:CallanScheme} presents Callan's perturbative renormalization scheme, including the $\theta$-insertion, the Callan--Symanzik equation, the finite formulation, and its relation to the functional flow equation. In section~\ref{sec:secondorder} we derive exact second-order flow equations, both as a one-loop equation involving the scale derivative of the inverse propagator and as a two-loop equation with one-loop and two-loop structures. Section~\ref{sec:secondaryFlowMethod} introduces the secondary flow method, in which derivatives with respect to an additional parameter $l$ yield ultraviolet-finite flow equations. Section~\ref{sec:ApplicationEffectivePotential} applies these methods to the effective potential of an $O(N)$-symmetric scalar theory in the local potential approximation, deriving the corresponding flow equations, solving the secondary flow in $d=2$, $3$, and $4$ dimensions, and discussing the flow of Taylor-expanded couplings and the physical significance of the integration constants. Section~\ref{sec:discussion} contains our discussion, conclusions, and outlook. Two appendices collect second-order flow equations for higher correlation functions and properties of the incomplete Gamma function.

\paragraph{Notation and conventions.}
We will occasionally use DeWitt's abstract index notation where e.\ g.\ $J_A \phi^A = \sum_j \int_x J_j(x) \phi_j(x)$. In other words, an uppercase abstract index $A=(x,j)$ combines a continuous position label $x$ or momentum label $p$ and a discrete index $j$, we use the Einstein summation convention, and the ``summation'' over $A$ includes an integral over continuous variables and a sum over $j$. The integral over spacetime or momentum is abbreviated as
\begin{equation}
  \int_x = \int d^d x, \quad\quad\quad \int_p = \int \frac{d^d p}{(2\pi)^d},
\end{equation}
where we work in $d$ Euclidean dimensions.

\section{First order functional flow equation}
\label{sec:FirstOrderFunctionalFlowEquation}

We first recall the standard, first order, one loop flow equation for the Schwinger functional, closely related to Polchinski's Wilsonian flow \cite{Polchinski_1984_Renormalization}, and for the effective action, due to Wetterich \cite{Wetterich_1993_Exact, Morris_1994_Exact}, and their derivation. This is standard material and included here for completeness and to establish our notation.

\subsection{Modified Schwinger functional and flow equation}
We consider a Schwinger functional,
\begin{equation}
  Z_{k}[J] = \exp\left(W_{k}[J] \right) = \int D\phi \, \exp\left(-S[\phi] - \Delta S_k[\phi] +  J_A \phi^A \right),
  \label{eq:ZWDefinition}
\end{equation}
where an infrared regulator term,
\begin{equation*}
  \Delta S_k[\phi]  = \frac{1}{2} R_{kAB} \phi^A \phi^B, 
\end{equation*}
has been added to the microscopic action $S[\phi]$. We are here particularly interested in the case where the matrix $R_{kAB} = k^2 \delta_{AB}$ has the form of an additional \textit{microscopic} mass squared term $k^2$, so that $\Delta S_k[\phi] = \int d^d x (1/2) k^2 \phi(x)^2$. In this case one can also see $W_k[J]$ as the Schwinger functional of a theory where the microscopic mass squared, encoded in $S[\phi]$, has been shifted.

A large part of the derivations below holds for general choice of $R_{kAB}$. In any case we assume that $R_{kAB}$ vanishes for $k=0$ so that the functional $W_k[J]$ approaches the standard Schwinger functional $W[J]$, the generating functional of connected correlation functions. The functional $W_k[J]$ is convex \cite{ORaifeartaigh_1986_Constraint} with respect to the argument $J$.

We shall also assume implicitly that there is a UV regularization present, such as for example a spacetime lattice or Schwinger proper time regularization. This can be seen as part of the definition of the functional integral measure $D\phi$.

Taking a derivative with respect to this infrared parameter, one finds the flow equation \cite{Polchinski_1984_Renormalization}
\begin{equation*}
  \frac{\partial}{\partial k} W_{k}[J] = - \frac{1}{2} \left(\frac{d}{dk} R_{kAB} \right) \left\{ (W_{k}^{(2)}[J])^{AB} + \frac{\delta W_k[J]}{\delta J_A} \frac{\delta W_k[J]}{\delta J_B} \right\}.
\label{eq:flowWFunctional}
\end{equation*}
On the right hand side we employ the field expectation value
\begin{equation}
  \frac{\delta W_k[J]}{\delta J_A} = \langle \phi^A \rangle = \Phi^A,
\end{equation}
and the connected two-point function,
\begin{equation}
(W_{k}^{(2)}[J])^{AB} = \frac{\delta^2 }{\delta J_A \delta J_B} W_{k}[J] = \left\langle \phi^A \phi^B \right\rangle - \Phi^A \Phi^B.
\end{equation}

\subsection{Modified quantum effective action}
One considers now the Legendre transform of the modified Schwinger functional $W_{k}[J]$,
\begin{equation}
  \tilde \Gamma_{k}[\Phi] = \sup_J \left( J_A \Phi^A - W_{k}[J] \right).
  \label{eq:defGammaTildeFunctional}
\end{equation}
We also define the so-called flowing action
\begin{equation}
  \Gamma_{k}[\Phi] = \tilde \Gamma_{k}[\Phi] - \frac{1}{2} R_{kAB} \Phi^A \Phi^B.
   \label{eq:solGammarkappa}
\end{equation}
For $k=0$, and therefore $R_{kAB}=0$, the definitions in \eqref{eq:defGammaTildeFunctional} and \eqref{eq:solGammarkappa} give the standard quantum effective action or one-particle irreducible effective action, $\tilde \Gamma_{0}[\Phi] = \Gamma_{0}[\Phi] = \Gamma[\Phi]$.

From the definition as a Legendre transform in eq.\ \eqref{eq:defGammaTildeFunctional} one finds the field equations
\begin{equation}
  (\tilde\Gamma_k^{(1)}[\Phi])_A = \frac{\delta}{\delta \Phi^A} \tilde\Gamma_k[\Phi] = J_A, \quad\quad\quad
  (\Gamma_k^{(1)}[\Phi])_A = \frac{\delta}{\delta \Phi^A} \Gamma_k[\Phi] = J_A - R_{kAB} \Phi^B,
\end{equation}
and relations among higher-order derivatives, like
\begin{equation}
	(\tilde \Gamma_k^{(2)}[\Phi])_{AB} (W^{(2)}_k[J])^{BC} = \delta_A^{\;\;C}, \quad\quad\quad
	(\Gamma_k^{(2)}[\Phi]+R_k)_{AB} (W^{(2)}_k[J])^{BC} = \delta_A^{\;\;C}.
\end{equation}

\subsection{Wetterich flow equation}\label{sec:WetterichFlowEquation}
From the definition \eqref{eq:defGammaTildeFunctional} and \eqref{eq:flowWFunctional} one obtains the flow equation \cite{Wetterich_1993_Exact, Morris_1994_Exact} with the parameter $k$,
\begin{equation}
  \frac{\partial}{\partial k} \tilde \Gamma_{k}[\Phi] =  \frac{1}{2} \left( \frac{d}{dk} R_{kAB} \right) \left\{  (\tilde \Gamma_{k}^{(2)}[\Phi]^{-1})^{AB} + \Phi^A \Phi^B \right\}.
\end{equation}
The solution to this functional differential equation is equivalent to a solution of the functional integral.

The flow equation for the flowing action $\Gamma_{k}[\Phi]$ reads
\begin{equation}
  \frac{\partial}{\partial k} \Gamma_k[\Phi] = \frac{1}{2} \left(\frac{d}{dk} R_{kAB} \right) ((\Gamma_{k}^{(2)}[\Phi] + R_k )^{-1})^{AB} = \frac{1}{2} \;
  \begin{tikzpicture}[baseline=(base)]
    \node (base) at (0,-0.6ex) {};
    \begin{feynman}
      \vertex (a1);
      \vertex[above=0.4cm of a1] (b1);
      \vertex[right=0.4cm of a1] (c1);
      \vertex[below=0.4cm of a1] (d1);
      \vertex[left=0.4cm of a1] (e1);
      \diagram*{
        (b1) -- [quarter left] (c1),
        (c1) -- [quarter left] (d1),
        (d1) -- [quarter left] (e1),
        (e1) -- [quarter left] (b1),
        };
        \draw[fill=black] (b1) circle [radius=0.08cm];
    \end{feynman}
  \end{tikzpicture}, 
\label{eq:flowequationGammafirst}
\end{equation}
where we have introduced a convenient graphical notation in the last step, with the dot representing the regulator derivative, and the line the full propagator.

Alternatively, one can write the flow equation as
\begin{equation}
  \frac{\partial}{\partial k} \Gamma_k[\Phi] = \frac{\tilde\partial}{\partial k} \frac{1}{2} \text{Tr} \left\{  \ln\left(\Gamma_{k}^{(2)}[\Phi] + R_k \right)\right\},
\end{equation}
with a formal derivative that acts only on $R_k$ on the right hand side. For the special case of a local mass-like regulator, $R_{kAB}=k^2 \delta_{AB}$, one may also write the flow equation in the alternative form
\begin{equation}
  \frac{\partial}{\partial k^2} \tilde \Gamma_k[\Phi] = \frac{\partial}{\partial l^2} \frac{1}{2} \left[ \text{Tr} \left\{  \ln\left(\tilde \Gamma_{k}^{(2)}[\Phi] + l^2 \right)\right\} + l^2 \delta_{AB} \Phi^A \Phi^B\right] {\Big |}_{l^2=0},
\end{equation}
a form that will be useful below to discuss a connection to Callan's perturbative renormalization scheme \cite{Alexandre_2001_Functional}.

For a local regulator of the type $R_{kAB} = k^2 \delta_{AB}$ this flow equation is for field theories usually not regular in the ultraviolet, or at small distances, because two-point correlation functions have distributional character. To avoid this one may introduce an additional UV regularization, for example the Schwinger proper time scheme \cite{Schwinger_1951_Gauge}, where one replaces the full propagator according to
\begin{equation}
  \tilde \Gamma_{k}^{(2)}[\Phi]^{-1} \to \int_{1/\Lambda^2}^\infty d\tau \, e^{-\tau \tilde \Gamma_{k}^{(2)}[\Phi]},
\end{equation}
where $\Lambda$ is a large momentum scale. Assuming positive real parts of the eigenvalues of $\tilde \Gamma^{(2)}[\Phi]$, so that the proper time integral converges at its upper boundary, one finds for $1/\Lambda^2 \to 0$ the original expression. For finite $\Lambda$, ultraviolet divergences are tamed\footnote{For discussions of the relation between proper time regularization and the renormalization group see \cite{Liao_1996_Connection, Litim_2001_Predictive, Litim_2002_Completeness, Bonanno_2005_Proper}.}.

\subsection{Flow equations for higher order correlation functions}
By taking functional derivatives of \eqref{eq:flowequationGammafirst} one obtains flow equations for functional derivatives of the flowing action \cite{Ellwanger_1993_Flow}, such as for the effective equation of motion,
\begin{equation}
  \begin{split}
  \frac{\partial}{\partial k}  (\Gamma^{(1)}_{k}[\Phi])_A = & \frac{\partial}{\partial k} \frac{\delta}{\delta \Phi^A} \Gamma_{k}[\Phi] =  \frac{\delta}{\delta \Phi^A} \frac{\partial}{\partial k} \Gamma_{k}[\Phi] \\ = &
  -\frac{1}{2} \left( \frac{d}{dk} R_{kBE} \right)((\Gamma_{k}^{(2)}[\Phi] + R_k)^{-1})^{BC} (\Gamma^{(3)}_{k}[\Phi])_{ACD} ((\Gamma_{k}^{(2)}[\Phi] + R_k)^{-1})^{DE},
  \end{split}
  \label{eq:flowequationGamma1first}
\end{equation}
and for the inverse propagator,
\begin{equation}
\begin{split}
  \frac{\partial}{\partial k} (\Gamma^{(2)}_{k}[\Phi])_{AB} = & \frac{\partial}{\partial k} \frac{\delta^2}{\delta \Phi^A \delta\Phi^B} \Gamma_{k}[\Phi] =  \frac{\delta^2}{\delta \Phi^A\delta \Phi^B} \frac{\partial}{\partial k} \Gamma_{k}[\Phi] \\
  = & \left( \frac{d}{dk} R_{kCH} \right) ((\Gamma_{k}^{(2)}[\Phi] + R_k )^{-1})^{CD} (\Gamma^{(3)}_{k}[\Phi])_{ADE} ((\Gamma_{k}^{(2)}[\Phi] + R_k)^{-1})^{EF} \\ & \quad\quad\quad\quad\quad \times (\Gamma^{(3)}_{k}[\Phi])_{BFG} ((\Gamma_{k}^{(2)}[\Phi] + R_k)^{-1})^{GH}  \\
  & -  \frac{1}{2} \left( \frac{d}{dk} R_{kCF} \right) ((\Gamma_{k}^{(2)}[\Phi] + R_k )^{-1})^{CD} (\Gamma^{(4)}_{k}[\Phi])_{ABDE} ((\Gamma_{k}^{(2)}[\Phi] + R_k )^{-1})^{EF}.
\end{split}\label{eq:flowequationGamma2first2}
\end{equation}
In this way one can go on. For $\partial \Gamma_{k}^{(n)}[\Phi]/\partial k$ one has $\Gamma_{k}^{(n+2)}$ as the highest power on the right hand side, so that the hierarchy of flow equations does not close \cite{Ellwanger_1993_Flow, Morris_1994_Exact, Berges_2002_Nonperturbative}.

These equations have a graphical representation as one-loop diagrams with one external leg,
\begin{equation}
  \frac{\partial}{\partial k}  (\Gamma^{(1)}_{k}[\Phi])_A = - \frac{1}{2}\;
  \begin{tikzpicture}[baseline=(base)]
    \node (base) at (0,-0.6ex) {};
    \begin{feynman}
      \vertex (a1);
      \vertex[above=0.4cm of a1] (b1);
      \vertex[right=0.4cm of a1] (c1);
      \vertex[below=0.4cm of a1] (d1);
      \vertex[left=0.4cm of a1] (e1);
      \vertex[below=0.55cm of a1] (f1) {\footnotesize \(A \)};
      \diagram*{
        (b1) -- [quarter left] (c1),
        (c1) -- [quarter left] (d1),
        (d1) -- [quarter left] (e1),
        (e1) -- [quarter left] (b1),
        (d1) -- (f1)
        };
        \draw[fill=black] (b1) circle [radius=0.08cm];
    \end{feynman}
  \end{tikzpicture}, 
\end{equation}
and two external legs, respectively,
\begin{equation}
  \frac{\partial}{\partial k}  (\Gamma^{(2)}_{k}[\Phi])_{AB} =
  \begin{tikzpicture}[baseline=(base)]
    \node (base) at (0,-0.6ex) {};
    \begin{feynman}
      \vertex (a1);
      \vertex[above=0.4cm of a1] (b1);
      \vertex[right=0.4cm of a1] (c1);
      \vertex[below=0.4cm of a1] (d1);
      \vertex[left=0.4cm of a1] (e1);
      \vertex[left=0.55cm of a1] (f1) {\footnotesize \(A \)};
      \vertex[right=0.55cm of a1] (g1) {\footnotesize \(B \)};
      \diagram*{
        (b1) -- [quarter left] (c1),
        (c1) -- [quarter left] (d1),
        (d1) -- [quarter left] (e1),
        (e1) -- [quarter left] (b1),
        (e1) -- (f1),
        (c1) -- (g1)
        };
      \draw[fill=black] (b1) circle [radius=0.08cm];
    \end{feynman}
  \end{tikzpicture} - \frac{1}{2} \;
  \begin{tikzpicture}[baseline=(base)]
    \node (base) at (0,-0.6ex) {};
    \begin{feynman}
      \vertex (a1);
      \vertex[above=0.4cm of a1] (b1);
      \vertex[right=0.4cm of a1] (c1);
      \vertex[below=0.4cm of a1] (d1);
      \vertex[left=0.4cm of a1] (e1);
      \vertex[below=0.6cm of a1] (f1);
      \vertex[left=0.15 cm of f1] (g1) {\footnotesize \(A \)};
      \vertex[right=0.15 cm of f1] (h1) {\footnotesize \(B \)};
      \diagram*{
        (b1) -- [quarter left] (c1),
        (c1) -- [quarter left] (d1),
        (d1) -- [quarter left] (e1),
        (e1) -- [quarter left] (b1),
        (d1) -- (g1),
        (d1) -- (h1)
        };
        \draw[fill=black] (b1) circle [radius=0.08cm];
    \end{feynman}
  \end{tikzpicture}. 
\label{eq:FlowGamma2FeynmanDiagram}
\end{equation}

\section{Callan's perturbative renormalization group scheme}
\label{sec:CallanScheme}

Callan's renormalization scheme, as discussed e.\ g.\ in ref.~\cite{Callan_1970_Broken, Symanzik_1970_Small, Blaer_1974_Field, Callan_1993_Introduction}, provides in the perturbative domain an alternative to the functional renormalization group approach reviewed in section \ref{sec:FirstOrderFunctionalFlowEquation}. A superficial difference is that  the Wetterich equation employs an infrared regulator $R_k(p)$ depending on the flow parameter $k$, Callan's scheme is organized around the physical squared mass as the renormalization group parameter. This difference can be minimized by choosing $R_k(p)$ independent of momentum $p$ \cite{Alexandre_2001_Functional}.

\subsection{The \texorpdfstring{$\theta$}{theta}-insertion}
Callan's scheme is formulated in perturbation theory with a microscopic or bare mass-squared $\bar m^2$ and a microscopic or bare coupling $\bar \lambda$, defined with an ultraviolet cutoff $\Lambda$. The central operation is the derivative with respect to $\bar m^2$ with other microscopic parameters fixed, like $\bar\lambda$ and $\Lambda$. In Euclidean conventions, differentiating an $n$-point vertex function $\Gamma^{(n)}$ with respect to $\bar m^2$ is equivalent to inserting the composite operator $\theta = \tfrac{1}{2}\phi^2$ at zero momentum into a diagram~\cite{Callan_1993_Introduction, Sirlin_1972_Mass},
\begin{equation}
  \frac{\partial}{\partial \bar m^2} \Gamma^{(n)} = \Gamma_\theta^{(n)}.
  \label{eq:thetaInsertion}
\end{equation}
The mechanism is simple: the derivative acts on each microscopic propagator as
\begin{equation}
  \frac{\partial}{\partial \bar m^2}\,\frac{1}{p^2 + \bar m^2} = -\frac{1}{(p^2 + \bar m^2)^2},
  \label{eq:propagatorDoubling}
\end{equation}
i.\,e.\ it replaces a single propagator by the negative of its square. This reduces the superficial degree of divergence of any Feynman diagram by two units. Repeated mass derivatives generate multiple $\theta$-insertions, like
\begin{equation}
  \Gamma_{\theta\theta}^{(n)} = \frac{\partial^2}{\partial (\bar m^2)^2}\Gamma^{(n)},
\end{equation}
and each additional insertion further reduces the degree of divergence by two.

\subsection{Field strength and \texorpdfstring{$\theta$}{theta}-renormalization and renormalization conditions}
One introduces a field renormalization factor $Z$, so that the renormalized fields are $\varphi(x) = \sqrt{Z} \Phi(x)$, where $\Phi(x) = \langle \phi(x) \rangle$ are the bare field expectation values, and
\begin{equation}
\bar \Gamma^{(n)}(x_1, \ldots, x_n) = \frac{\delta^n}{\delta \varphi(x_1) \ldots \varphi(x_n)} \Gamma[\varphi] = \frac{1}{Z^{n/2}} \frac{\delta^n}{\delta \Phi(x_1) \ldots \delta \Phi(x_n)}\Gamma[\Phi] = \frac{1}{Z^{n/2}} \Gamma^{(n)}(x_1, \ldots, x_n),
\end{equation}
are the corresponding renormalized vertex functions. The usual convention is to fix $Z$ such that the kinetic term in the quantum effective action has standard normalization after renormalization, i.~e.~
\begin{equation}
  \frac{d\bar \Gamma^{(2)}(p)}{dp^2} = 1,
\end{equation}
either at vanishing momentum $p=0$ or on-shell, at the zero-crossing of $\bar \Gamma^{(2)}(p)$ corresponding to a freely propagating particle.

Callan also introduced an additional renormalization factor $Z_\theta$ for the $\theta$-insertion, such that its renormalized form is
\begin{equation}
\bar \Gamma_\theta^{(n)}(x_1, \ldots, x_n) = \frac{1}{Z_\theta Z^{n/2}} \Gamma_\theta^{(n)}(x_1, \ldots, x_n).
\label{eq:ZthetaRenormlizationCorrelationFunction}
\end{equation}
The need for an additional renormalization constant $Z_\theta$ can be understood from the point of view of the general theory of composite operator renormalization \cite{Keller_1992_Perturbative, Keller_1993_Perturbative}. It can for example be fixed such that the renormalized $\theta$ insertion is intuitively a derivative with respect to the full renormalized mass-squared instead of the microscopic one. Callan consequently proposed to fix $Z_\theta$ such that $\bar \Gamma_\theta^{(2)}(p) = 1$, at vanishing momentum $p=0$, but this choice is not strictly necessary.

One defines the renormalized mass parameter $m$ either through
\begin{equation}
\bar \Gamma^{(2)}(0) = m^2,
\label{eq:RGConditionm2}
\end{equation}
or the on-shell condition $\bar \Gamma^{(2)}(p){\big |}_{p^2+m^2=0}=0$ for the renormalized inverse propagator. The renormalized coupling $\lambda$ is defined similarly with $\bar\Gamma^{(4)} = \lambda$ at vanishing momentum, or on-shell.

\subsection{Callan--Symanzik equation}

To replace all bare quantities by renormalized ones, one uses that correlation functions in perturbation theory depend only on the parameters $m^2$ and $\lambda$. This allows to rewrite derivatives acting on correlation functions,
\begin{equation}
	\frac{\partial}{\partial \bar m^2} = \frac{d m^2}{d \bar m^2} \frac{\partial}{\partial m^2} + \frac{d \lambda}{d \bar m^2} \frac{\partial}{\partial \lambda}.
\label{eq:massDerivativeDistributed}
\end{equation}
Multiplying with $2m^2$ and dividing by $d m^2 /d \bar m^2$ leads to the Callan--Symanzik equations~\cite{Callan_1970_Broken, Symanzik_1970_Small, Callan_1993_Introduction} (see also \cite{Amit_1984_Field, Costello_2011_Renormalization, Collins_2023_Renormalization, Zinn-Justin_2021_Quantum} for textbook accounts) for renormalized vertex functions,
\begin{equation}
\begin{split}
\left[ 2m^2 \frac{\partial}{\partial m^2} + \beta \frac{\partial}{\partial \lambda} - \frac{\eta n}{2} \right] \bar\Gamma^{(n)} = & 2m^2 G \, \bar\Gamma_\theta^{(n)},\\
\left[ 2m^2 \frac{\partial}{\partial m^2} + \beta \frac{\partial}{\partial \lambda} - \frac{\eta n}{2} + \eta_\theta \right] \bar \Gamma_\theta^{(n)} = & 2m^2 G \, \bar\Gamma_{\theta\theta}^{(n)}.
\end{split}
\label{eq:CallanSymanzikCorrelationFunction01}
\end{equation}
One may equivalently express these in terms of generating functionals,
\begin{equation}
\begin{split}
\left[ 2m^2 \frac{\partial}{\partial m^2} + \beta \frac{\partial}{\partial \lambda} - \frac{\eta }{2} \int d^d x \varphi(x) \frac{\delta}{\delta \varphi(x)} \right] \bar \Gamma[\varphi] = & 2m^2 G \, \bar \Gamma_\theta[\varphi], \\
\left[ 2m^2 \frac{\partial}{\partial m^2} + \beta \frac{\partial}{\partial \lambda} - \frac{\eta }{2} \int d^d x \varphi(x) \frac{\delta}{\delta \varphi(x)} + \eta_\theta \right] \bar \Gamma_\theta[\varphi] = & 2m^2 G \, \bar \Gamma_{\theta\theta}[\varphi],
\end{split}\label{eq:CallanSymanzik}
\end{equation}
for the perturbative contribution to the effective action $\Gamma[\varphi]$, $\Gamma_\theta[\varphi]$ and $\Gamma_{\theta\theta}[\varphi]$ with zero, one or two $\theta$-insertions. The right-hand side of the first equation is a generating functional for vertex functions with one additional $\theta$-insertion. It satisfies its own Callan--Symanzik equation with an additional $\theta$-insertion, and so one could go on. For formulations where this tower of equations is further unified into a single equation, see refs.\ \cite{MacDowell_1975_Generalized, Ageeva_2025_Nonrenormalizable}.

We use here the abbreviations\footnote{Note that some authors, including Callan, work with $\gamma=-\eta/2$ instead.}
\begin{equation}
\begin{split}
\beta = &  2 m^2 \frac{1}{\frac{d m^2}{d \bar m^2}} \frac{d \lambda}{d \bar m^2} 
= m \frac{d\lambda}{d m},\\
\eta = & - \frac{2m^2}{Z\frac{d m^2}{d \bar m^2}}\frac{d Z}{d \bar m^2} = - \frac{m}{Z} \frac{dZ}{dm}, \\
\eta_\theta = & - \frac{2m^2}{Z_\theta \frac{d m^2}{d \bar m^2}}\frac{d Z_\theta }{d \bar m^2} = - \frac{m}{Z_\theta} \frac{dZ_\theta}{dm}, \\
G = & \frac{Z_\theta}{\frac{d m^2}{d \bar m^2}},
\end{split}\label{eq:AbbreviationsCSE}
\end{equation}

Evaluating eq.\ \eqref{eq:CallanSymanzikCorrelationFunction01} for $n=2$, and using the conditions \eqref{eq:RGConditionm2} and Callan's additional condition $\bar \Gamma_\theta^{(2)}(0)=1$ gives the relation
\begin{equation}
	G = 1-\frac{\eta}{2}.
\label{eq:AdditionalRelationForGCallan}
\end{equation}

\subsection{Finite formulation}
The observation that a sufficient number of $\theta$-operations renders perturbative Feynman diagrams finite is the basis of the \emph{finite formulation} of quantum field theory, originating in early work \cite{Callan_1993_Introduction, Sirlin_1972_Mass} and revived and further developed by Mooij and Shaposhnikov~\cite{Mooij_2023_QFT,Mooij_2023_Finite}\footnote{A related finite renormalization scheme has been proposed in ref.\ \cite{THooft_2005_Renormalization}.}. In $d$ spacetime dimensions, a one-loop integral with $n$ propagators converges in the ultraviolet when $n > d/2$. Since each $\theta$-insertion adds one propagator, one can always choose enough insertions to achieve convergence. In $d=4$ scalar $\phi^4$ theory, two $\theta$-operations suffice for the two-point function, one for the four-point function, and diagrams with $n > 4$ external legs are already convergent~\cite{Mooij_2023_QFT}.

Within this framework, the Callan--Symanzik equations for $\theta$-inserted vertex functions, together with the renormalization conditions, allow the computation of all quantum corrections order by order in the coupling without encountering intermediate divergences. The method bypasses the need for regularization and counterterms: all Feynman integrals that appear are manifestly convergent.

The effective potential has been studied within this framework by Mooij and Shaposhnikov~\cite{Mooij_2024_Effective}, who derived the Callan--Symanzik equation for the effective potential and its mass derivatives and showed that the quantum corrections can be expressed in terms of a finite set of parameters fixed by renormalization conditions.

The method has been extended to nonrenormalizable theories by Ageeva, Petrov, and Shaposhnikov~\cite{Ageeva_2025_Nonrenormalizable}. Here one has some interesting modifications of the scheme: for a calculation to all orders in perturbation theory one would in principle need infinitely many $\theta$-insertions to arrive at finite expressions, and corresponding boundary terms to solve the equations. However, for a calculation at a fixed order in perturbation theory one needs only a finite number of $\theta$ insertions, and the scheme becomes useful.

\subsection{Relation to the functional flow equation}\label{sec:RelationCallanToFunctionalFormulation}

Callan's renormalization scheme is certainly similar in spirit to the functional renormalization group in the formulation for the quantum effective action, and a mass-like regulator \cite{Alexandre_2001_Functional, Alexandre_2002_Functional}. More precisely, the correlation functions $\Gamma^{(n)}$ are functional derivatives of $\tilde \Gamma_k[\Phi]$, the functional introduced in eq.\ \eqref{eq:defGammaTildeFunctional},
\begin{equation}
\Gamma^{(n)}(x_1, \ldots, x_n) = \frac{\delta^n}{\delta \Phi(x_1) \ldots \delta \Phi(x_n)} \tilde\Gamma_k[\Phi],
\end{equation}
where $\Phi(x)=\langle \phi(x) \rangle$ are expectation values of bare fields, and the regulator implements a shift in the microscopic mass-squared,
\begin{equation}
	R_k = \Delta \bar m^2.
\label{eq:Deltam2Regulator}
\end{equation}
Similarly, $\Gamma_\theta^{(n)}$ are correlation functions with an additional derivative with respect to $\Delta \bar m^2$,
\begin{equation}
	\Gamma^{(n)}_\theta(x_1, \ldots, x_n) = \frac{\delta^n}{\delta \Phi(x_1) \ldots \delta \Phi(x_n)} \frac{\partial}{\partial \Delta \bar m^2}\tilde\Gamma_k[\Phi].
\end{equation}
The term on the right can in principle be evaluated with the Wetterich equation,
\begin{equation}
\begin{split}
	\frac{\partial}{\partial \Delta \bar m^2} \tilde \Gamma_k[\Phi] 
	= & \frac{1}{2} \left[ \text{Tr}\left\{ \tilde \Gamma^{(2)}_k[\Phi]^{-1} \right\} + \int d^d x \, \Phi(x)^2 \right],
\end{split}\label{eq:flowAlternativeWithL}
\end{equation}
for the regulator \eqref{eq:Deltam2Regulator}.

However, this is not what is done in Callan's scheme. There $\Gamma_\theta^{(n)}$ are rather understood as derivatives of \textit{perturbative} contributions to an effective vertex, calculated to a given order in perturbation theory, with respect to the microscopic squared mass $\bar m^2$.

The change from bare field expectation values $\Phi(x)$ to rescaled or renormalized fields $\varphi(x) = \sqrt{Z} \Phi(x)$ is straight-forward and done also for the study of functional renormalization group equations. We write the effective vertices for renormalized fields as
\begin{equation}
	\bar \Gamma^{(n)}(x_1, \ldots, x_n) = \frac{\delta^n}{\delta \varphi(x_1) \ldots \varphi(x_n)} \tilde \Gamma_k[\Phi] = \frac{1}{Z^{n/2}} \frac{\delta^n}{\delta \Phi(x_1) \ldots \delta \Phi(x_n)}\tilde \Gamma_k[\Phi] = \frac{1}{Z^{n/2}} \Gamma^{(n)}(x_1, \ldots, x_n).
\end{equation}

More interesting is the additional renormalization of the $\theta$-insertion by $Z_\theta$ in eq.\ \eqref{eq:ZthetaRenormlizationCorrelationFunction}.

We may understand this as a reparametrization of the regulator,
\begin{equation}
R_k =	\Delta \bar m^2 = \Delta \bar m^2(k^2), \quad\quad\quad \frac{1}{Z_\theta} = \frac{d\Delta \bar m^2}{d k^2} = \frac{d R_k}{dk^2}, \quad\quad\quad \frac{\partial}{\partial k^2} = \frac{d \Delta \bar m^2}{d k^2}\frac{\partial}{\partial \Delta \bar m^2} = \frac{1}{Z_\theta} \frac{\partial}{\partial \Delta \bar m^2}.
\label{eq:flowVelocityChange}
\end{equation}
In any case the regulator is independent of momentum, and such a reparametrization just changes the ``flow velocity''. If all flow equations were finite, it could for example be done such that the renormalized squared mass, fixed through the condition
\begin{equation}
	\bar \Gamma^{(2)}(p) {\big |}_{p=0} = \tilde \Gamma^{(2)}_k(p) {\big |}_{p=0} = m^2,
\end{equation}
changes with ``flow velocity''
\begin{equation}
	\frac{d m^2}{d k^2} = \frac{1}{1-\frac{\eta}{2}}.
\label{eq:flowVelocityChoiceCallan}
\end{equation}
This would correspond to Callan's choice of $Z_\theta$. On the other side, one would expect that any physical result is independent of the choice of the flow parametrization or ``flow velocity'', and we will see below that this is indeed the case.

With this we can write the renormalized correlation functions with $\theta$-insertions as
\begin{equation}
\begin{split}
\bar \Gamma^{(n)}_\theta(x_1, \ldots, x_n) & =  \frac{\delta^n}{\delta \varphi(x_1)\ldots \delta \varphi(x_n)} \bar \Gamma_\theta[\varphi] = \frac{\delta^n}{\delta \varphi(x_1)\ldots \delta \varphi(x_n)} \frac{\partial}{\partial k^2} \tilde \Gamma_k[\Phi] \\
& = \frac{1}{Z^{n/2}} \frac{\delta^n}{\delta \Phi(x_1)\ldots \delta \Phi(x_n)} \frac{\partial}{\partial k^2} \tilde \Gamma_k[\Phi].
\end{split}
\end{equation}
Note that the derivative with respect to $k^2$ is done at fixed $\Phi$ and not $\varphi$. Of course one could still use a variant of the Wetterich equation for the right hand side,
\begin{equation}
	\bar \Gamma_\theta[\varphi] = \frac{\partial}{\partial k^2} \tilde \Gamma_k[\Phi] =  \frac{1}{2} \frac{d R_k}{dk^2} \left[ \text{Tr}\left\{ \tilde \Gamma^{(2)}_k[\Phi]^{-1} \right\} + \int d^d x \, \Phi(x)^2 \right],
\end{equation}
still with momentum-independent regulator $R_k$.

A flow equation where the renormalized fields $\varphi(x) = \sqrt{Z} \Phi(x)$ are kept fixed instead of the bare field expectation values $\Phi$ can be obtained with the chain rule,
\begin{equation}
\begin{split}
	\frac{\partial}{\partial k^2} \tilde \Gamma_k {\big |}_{\Phi} = & \frac{\partial}{\partial k^2} \tilde \Gamma_k {\big |}_{\varphi} + \int d^d x  \left( \frac{\delta}{\delta \varphi(x)} \tilde \Gamma_k \right) \frac{\partial}{\partial k^2} \varphi(x) {\big |}_{\Phi(x)} \\
	= & \frac{\partial}{\partial k^2} \tilde \Gamma_k {\big |}_{\varphi} + \frac{1}{2Z}\frac{dZ}{dk^2} \int d^d x  \, \varphi(x) \frac{\delta}{\delta \varphi(x)} \tilde \Gamma_k
\end{split}
\end{equation}
We arrive at the intermediate result
\begin{equation}
\begin{split}
	& \frac{\partial}{\partial k^2} \tilde \Gamma_k[\varphi] + \frac{1}{2Z}\frac{dZ}{dk^2} \int d^d x  \, \varphi(x) \frac{\delta}{\delta \varphi(x)} \tilde \Gamma_k[\varphi] \\
	& = \bar \Gamma_\theta[\varphi] = \frac{1}{2 Z} \frac{d R_k}{dk^2} \left[ \text{Tr}\left\{ \tilde \Gamma^{(2)}_k[\varphi]^{-1} \right\} + \int d^d x \, \varphi(x)^2 \right],
\end{split}
\end{equation}
where $\tilde \Gamma^{(2)}_k[\varphi]$ is now the second functional derivative of $\tilde \Gamma_k[\varphi] = \bar \Gamma[\varphi]$ with respect to the renormalized field $\varphi(x)$.

In a non-perturbative, functional formulation the effective action $\tilde \Gamma_k[\phi]$ depends on a renormalized squared mass $m^2$, and (infinitely) many parameters that we may denote as $\lambda_j$. One can rewrite the derivative $\partial/\partial k^2$ acting on $\tilde \Gamma_k[\varphi]$ in generalization of eq.\ \eqref{eq:massDerivativeDistributed} as
\begin{equation}
	\frac{\partial}{\partial k^2} = \frac{d m^2}{d k^2} \frac{\partial}{\partial m^2} + \sum_j \frac{d \lambda_j}{d k^2} \frac{\partial}{\partial \lambda_j},
\end{equation}
and after multiplication by $2m^2$ and dividing by $d m^2/d k^2$ we arrive at
\begin{equation}
	\left[ m \frac{\partial}{\partial m} + \sum_j \beta_j \frac{\partial}{\partial \lambda_j} - \frac{\eta}{2} \int d^d x \, \varphi(x) \frac{\delta}{\delta \varphi(x)} \right] \tilde \Gamma_k[\varphi] = 2 m^2 G \, \bar \Gamma_\theta[\varphi].
	\label{eq:CallanSymanzikNonPerturbative}
\end{equation}
We use here the abbreviations
\begin{equation}
\begin{split}
\beta_j = & 2 m^2 \frac{1}{\frac{d m^2}{d k^2}} \frac{d\lambda_j}{d k^2} = m \frac{d\lambda_j}{dm}, \\
  \eta = & - \frac{2m^2}{Z\frac{dm^2}{d k^2}}\frac{d Z}{d k^2} = - \frac{m}{Z} \frac{dZ}{dm}, \\
  G = &  \frac{1}{\frac{dm^2}{dk^2}}.
\end{split}
\end{equation}
Note that these are in agreement with the ones in eq.\ \eqref{eq:AbbreviationsCSE} . Note that also the expression for $G$ agrees if one takes into account \eqref{eq:flowVelocityChange}. Moreover, if the ``flow velocity'' is chosen such that eq.\ \eqref{eq:flowVelocityChoiceCallan} is obeyed, also eq.\ \eqref{eq:AdditionalRelationForGCallan} is fulfilled.

In Callan's perturbative renormalization scheme one would write down perturbative expressions for $\bar \Gamma_\theta[\varphi]$ to a given order in perturbation theory. Instead, from the non-perturbative functional renormalization group point of view eq.\ \eqref{eq:CallanSymanzikNonPerturbative} is supplemented with the Wetterich-type flow equation
\begin{equation}
	\bar \Gamma_\theta[\varphi] = \frac{1}{2 Z} \frac{d R_k}{dk^2} \left[ \text{Tr}\left\{ \tilde \Gamma^{(2)}_k[\varphi]^{-1} \right\} + \int d^d x \, \varphi(x)^2 \right].
\end{equation}
Formally, both the left hand side and the right hand side of eq.\ \eqref{eq:CallanSymanzikNonPerturbative} are invariant under redefinitions of the flow parameter $k$, or changes in the ``flow velocity''. A standard choice would be $R_k = Z k^2$, such that
\begin{equation}
 \frac{1}{Z} \frac{d R_k}{dk^2} = 1 + \frac{k}{2Z} \frac{dZ}{dk}.
\end{equation}

So far we assumed implicitly that $dm^2/dk^2$ is finite. It can happen, for example for scalar $\phi^4$ theory in $d=4$ dimensions, and for a regular choice of $R_k$, that $dm^2/dk^2$ is strongly UV dominated and in fact logarithmically UV divergent \cite{Alexandre_2001_Functional}. This would make $G$ and therefore the flow in Eq.\ \eqref{eq:CallanSymanzikNonPerturbative} ill-defined.

While Callan's method allows to take further derivatives with respect to $\bar m^2$, equivalent to further $\theta$-insertions, to arrive at UV finite expressions, we have so far discussed only one first order flow equation in section \ref{sec:WetterichFlowEquation}. This consideration motivates to study higher order flow equations, in order to see whether one can arrive at finite, non-perturbative flow equations with a mass-type infrared regulator, similar to Callan's perturbative scheme.

\section{Second order, one and two loop flow equations}
\label{sec:secondorder}
In this section we derive a new form of flow equation for the effective action that is formally a second order functional differential equation and has a two-loop structure\footnote{We were made aware by G.~P.~Vacca that equations of this kind were also derived and investigated in the diploma thesis of L.~Rossi \cite{Rossi_2019_Functional}.}. We are motivated by Callan's perturbative renormalization scheme, where higher order derivatives appear naturally, as well. We will base the discussion on the flowing action $\Gamma_k[\Phi]$. For generality we will keep $R_{kAB}$ open as a matrix. We use $k$ as the primary flow parameter, but one could easily adapt the formulas to other convenient flow parameters, like $k^2$ or the microscopic squared mass.

\subsection{Second order one-loop flow equation}
By taking another derivative on both sides of eq. \eqref{eq:flowequationGammafirst} with respect to the regularization parameter $k$, we obtain a flow equation of second order,
\begin{equation}
\begin{split}
  \frac{\partial^2}{\partial k^2} \Gamma_{k}[\Phi] = & \frac{1}{2} \left(\frac{d^2}{dk^2} R_{kAB} \right) ((\Gamma_{k}^{(2)}[\Phi] + R_k )^{-1})^{AB} \\
  &  -\frac{1}{2} \left( \frac{d}{dk} R_{kCD} \right) ((\Gamma_{k}^{(2)}[\Phi] + R_k )^{-1})^{CA} \left( \frac{d}{dk} R_{kAB} \right) ((\Gamma_{k}^{(2)}[\Phi] + R_k )^{-1})^{BD}\\
  &  -\frac{1}{2} \left( \frac{d}{dk} R_{kCD} \right) ((\Gamma_{k}^{(2)}[\Phi] + R_k )^{-1})^{CA} \left( \frac{\partial}{\partial k} (\Gamma^{(2)}_{k}[\Phi])_{AB} \right) ((\Gamma_{k}^{(2)}[\Phi] + R_k )^{-1})^{BD}.
\end{split}
\label{eq:flowequationGamma2first}
\end{equation}
Note that this contains the first order flow equation for the inverse propagator in the last term.

Diagrammatically, we have now a one-loop expression,
\begin{equation}
  \frac{\partial^2}{\partial k^2} \Gamma_k[\Phi] = \frac{1}{2}
  \;
  \begin{tikzpicture}[baseline=(base)]
    \node (base) at (0,-0.6ex) {};
    \begin{feynman}
      \vertex (a1);
      \vertex[above=0.4cm of a1] (b1);
      \vertex[right=0.4cm of a1] (c1);
      \vertex[below=0.4cm of a1] (d1);
      \vertex[left=0.4cm of a1] (e1);
      \diagram*{
        (b1) -- [quarter left] (c1),
        (c1) -- [quarter left] (d1),
        (d1) -- [quarter left] (e1),
        (e1) -- [quarter left] (b1),
        };
        \draw[fill=white] (b1) circle [radius=0.08cm];
    \end{feynman}
  \end{tikzpicture} \;
    - \frac{1}{2} \;
  \begin{tikzpicture}[baseline=(base)]
    \node (base) at (0,-0.6ex) {};
    \begin{feynman}
      \vertex (a1);
      \vertex[above=0.4cm of a1] (b1);
      \vertex[right=0.4cm of a1] (c1);
      \vertex[below=0.4cm of a1] (d1);
      \vertex[left=0.4cm of a1] (e1);
      \diagram*{
        (b1) -- [quarter left] (c1),
        (c1) -- [quarter left] (d1),
        (d1) -- [quarter left] (e1),
        (e1) -- [quarter left] (b1),
        };
      \draw[fill=black] (b1) circle [radius=0.08cm];
      \draw[fill=black] (d1) circle [radius=0.08cm];
    \end{feynman}
  \end{tikzpicture} \; - \frac{1}{2} \;
\begin{tikzpicture}[baseline=(base)]
  \node (base) at (0,-0.6ex) {};
  \begin{feynman}
    \vertex (a1);
    \vertex[above=0.4cm of a1] (b1);
    \vertex[right=0.4cm of a1] (c1);
    \vertex[below=0.4cm of a1] (d1);
    \vertex[left=0.4cm of a1] (e1);
    \diagram*{
      (b1) -- [quarter left] (c1),
      (c1) -- [quarter left] (d1),
      (d1) -- [quarter left] (e1),
      (e1) -- [quarter left] (b1),
      };
    \draw[fill=black] (b1) circle [radius=0.08cm];
    \draw[fill=white] ($(d1)+(-0.07cm,-0.07cm)$) rectangle ($(d1)+(0.07cm,0.07cm)$);
  \end{feynman}
\end{tikzpicture}
\end{equation}
with a single regulator derivative $dR_k/dk$ marked by a filled dot, a second regulator derivative $d^2 R_k / dk^2$ marked by an open dot, and the first derivative of the inverse propagator $\frac{\partial}{\partial k} \Gamma_k^{(2)}[\Phi]$ marked by an open square.

We remark that the first term in eq.\ \eqref{eq:flowequationGamma2first} would disappear for $R_k=k^2$ if $k^2$ is used as a flow parameter instead of $k$. The second term has indeed improved convergence properties in the ultraviolet, because it contains two cutoff insertions. There might sometimes be a remaining UV divergence, but if it concerns only field-independent terms, it is harmless and can easily be removed by an additive renormalization.

Finally, the last term in eq.\ \eqref{eq:flowequationGamma2first} is proportional to $\frac{\partial}{\partial k}\Gamma_k^{(2)}[\Phi]$. In the ultraviolet limit of very large momenta one can often expect this scale derivative of the full inverse propagator to vanish on physical grounds. In that case one would obtain a fully UV finite flow equation, featuring first and second order derivatives with respect to the scale parameter.

Here we have two possible options:
\begin{enumerate}
\item We could work with a setup where $\Gamma_k[\Phi]$, as well as the first scale derivative $\frac{\partial}{\partial k} \Gamma_k[\Phi]$ of the inverse propagator are being provided at some scale $k$, and \eqref{eq:flowequationGamma2first} is used as a differential equation to determine $\Gamma_k[\Phi]$ at all scales. This allows to devise truncation schemes that parametrize not only $\Gamma_k[\Phi]$, but also $\frac{\partial}{\partial k} \Gamma_k[\Phi]$ (or actually $\frac{\partial}{\partial k}\Gamma_k^{(2)}[\Phi]$).
\item We could insert the flow equation \eqref{eq:flowequationGamma2first2} for the inverse propagator in eq.\ \eqref{eq:flowequationGamma2first}, to arrive at a second order flow equation where the right hand side is fully determined by $\Gamma_k[\Phi]$.
\end{enumerate}

\subsection{Second order two-loop flow equation}
Following the second option, inserting the first-order flow equation for the inverse propagator leads to
\begin{equation}
\begin{split}
  \frac{\partial^2}{\partial k^2} \Gamma_{k}[\Phi] = & \frac{1}{2} \left(\frac{d^2}{dk^2} R_{kAB} \right) ((\Gamma_{k}^{(2)}[\Phi] + R_k )^{-1})^{AB} \\
  & - \frac{1}{2} \left( \frac{d}{dk} R_{kCF} \right) \left(\frac{d}{dk} R_{kDE} \right) ((\Gamma_{k}^{(2)}[\Phi] + R_k )^{-1})^{CD} ((\Gamma_{k}^{(2)}[\Phi] + R_k )^{-1})^{EF} \\
  & + \frac{1}{4} \left( \frac{d}{dk} R_{kIJ} \right) \left( \frac{d}{dk} R_{kCF} \right) ((\Gamma_{k}^{(2)}[\Phi] + R_k )^{-1})^{IA} ((\Gamma_{k}^{(2)}[\Phi] + R_k )^{-1})^{CD} \\ & \quad\quad\quad\quad\quad \times (\Gamma^{(4)}_{k}[\Phi])_{ABDE} ((\Gamma_{k}^{(2)}[\Phi] + R_k )^{-1})^{EF} ((\Gamma_{k}^{(2)}[\Phi] + R_k )^{-1})^{BJ} \\
  &  -\frac{1}{2} \left( \frac{d}{dk} R_{kIJ} \right) \left( \frac{d}{dk} R_{kCH} \right) ((\Gamma_{k}^{(2)}[\Phi] + R_k )^{-1})^{IA}
  ((\Gamma_{k}^{(2)}[\Phi] + R_k)^{-1})^{CD} \\ & \quad\quad\quad\quad\quad \times(\Gamma^{(3)}_{k}[\Phi])_{ADE} ((\Gamma_{k}^{(2)}[\Phi] + R_k)^{-1})^{EF} (\Gamma^{(3)}_{k}[\Phi])_{BFG} \\ & \quad\quad\quad\quad\quad \times  ((\Gamma_{k}^{(2)}[\Phi] + R_k)^{-1})^{GH}
  ((\Gamma_{k}^{(2)}[\Phi] + R_k )^{-1})^{BJ} .
\end{split}\label{eq:flowequationGammaSecond}
\end{equation}
Diagrammatically, we have now a mix of one- and two-loop terms,
\begin{equation}
  \frac{\partial^2}{\partial k^2} \Gamma_k[\Phi] = \frac{1}{2}
  \;
  \begin{tikzpicture}[baseline=(base)]
    \node (base) at (0,-0.6ex) {};
    \begin{feynman}
      \vertex (a1);
      \vertex[above=0.4cm of a1] (b1);
      \vertex[right=0.4cm of a1] (c1);
      \vertex[below=0.4cm of a1] (d1);
      \vertex[left=0.4cm of a1] (e1);
      \diagram*{
        (b1) -- [quarter left] (c1),
        (c1) -- [quarter left] (d1),
        (d1) -- [quarter left] (e1),
        (e1) -- [quarter left] (b1),
        };
        \draw[fill=white] (b1) circle [radius=0.08cm];
    \end{feynman}
  \end{tikzpicture} \;
    - \frac{1}{2} \;
  \begin{tikzpicture}[baseline=(base)]
    \node (base) at (0,-0.6ex) {};
    \begin{feynman}
      \vertex (a1);
      \vertex[above=0.4cm of a1] (b1);
      \vertex[right=0.4cm of a1] (c1);
      \vertex[below=0.4cm of a1] (d1);
      \vertex[left=0.4cm of a1] (e1);
      \diagram*{
        (b1) -- [quarter left] (c1),
        (c1) -- [quarter left] (d1),
        (d1) -- [quarter left] (e1),
        (e1) -- [quarter left] (b1),
        };
      \draw[fill=black] (b1) circle [radius=0.08cm];
      \draw[fill=black] (d1) circle [radius=0.08cm];
    \end{feynman}
  \end{tikzpicture} \; + \frac{1}{4} \;
  \begin{tikzpicture}[baseline=(base)]
    \node (base) at (0,-0.6ex) {};
    \begin{feynman}
      \vertex (a1);
      \vertex[right=0.2cm of a1] (f1);
      \vertex[left=0.2cm of a1] (g1);
      \vertex[above=0.2cm of f1] (f2);
      \vertex[below=0.2cm of f1] (f3);
      \vertex[above=0.2cm of g1] (g2);
      \vertex[below=0.2cm of g1] (g3);
      \vertex[above=0.4cm of a1] (b1);
      \vertex[right=0.4cm of a1] (c1);
      \vertex[below=0.4cm of a1] (d1);
      \vertex[left=0.4cm of a1] (e1);
      \diagram*{
        (b1) -- [quarter left] (f2),
        (f2) -- [quarter left] (a1),
        (a1) -- [quarter left] (g2),
        (g2) -- [quarter left] (b1),
        (a1) -- [quarter left] (f3),
        (f3) -- [quarter left] (d1),
        (d1) -- [quarter left] (g3),
        (g3) -- [quarter left] (a1)
        };
      \draw[fill=black] (b1) circle [radius=0.08cm];
      \draw[fill=black] (d1) circle [radius=0.08cm];
    \end{feynman}
  \end{tikzpicture} \; - \frac{1}{2} \;
  \begin{tikzpicture}[baseline=(base)]
    \node (base) at (0,-0.6ex) {};
    \begin{feynman}
      \vertex (a1);
      \vertex[above=0.4cm of a1] (b1);
      \vertex[right=0.4cm of a1] (c1);
      \vertex[below=0.4cm of a1] (d1);
      \vertex[left=0.4cm of a1] (e1);
      \diagram*{
        (b1) -- [quarter left] (c1),
        (c1) -- [quarter left] (d1),
        (d1) -- [quarter left] (e1),
        (e1) -- [quarter left] (b1),
        (c1) -- (e1)
        };
      \draw[fill=black] (b1) circle [radius=0.08cm];
      \draw[fill=black] (d1) circle [radius=0.08cm];
    \end{feynman}
  \end{tikzpicture} \, ,
\label{eq:SecondOrderFlowEqFeynman}
\end{equation}
with each loop containing a regulator derivative, $dR_k/dk$, marked by a filled dot, or a second regulator derivative, $d^2 R_k / dk^2$, marked by an open dot.

By taking functional derivatives with respect to $\Phi$ one finds two-loop flow equations for higher correlation functions, some of which are presented in appendix \ref{app:HigherDerivatives}.

Having a flow equation with two-loop terms could have many advantages for practical calculations \cite{Papenbrock_1995_Two-loop, Litim_2002_Perturbation, Carrington_2019_Renormalization, Blaizot_2021_Functional}, for example for effects that are known from perturbative investigations to rely on a two-loop structure.

As beautiful as they are, what the flow equations \eqref{eq:flowequationGamma2first} and \eqref{eq:flowequationGammaSecond} are in general not doing is to fully solve the ultraviolet convergence issues. The topology of the second (one-loop) diagram in \eqref{eq:SecondOrderFlowEqFeynman} looks promising in that regard, but the third term (two-loop) still leads to subdiagrams with logarithmic divergences \cite{Collins_2023_Renormalization}, e.\ g.\  for a scalar quantum field theory in $d=4$ dimensions, in the local potential approximation.

More generally, one can convince oneself that also for flow equations of higher orders, like $\partial^3 \Gamma_k[\Phi]/\partial k^3$, which contains three-loop terms, one always finds subdiagrams with the topology of a tadpole, like the last term in eq.\ \eqref{eq:FlowGamma2FeynmanDiagram}. Those are still logarithmically ultraviolet divergent in $d=4$ dimensions.

\section{Secondary flow method}
\label{sec:secondaryFlowMethod}

\subsection{Generalizing the flow equation kernel}
We now follow a different strategy to derive a higher-order flow equation. Here the idea is to modify the right hand side of the standard, first order flow equation by hand, and to determine it by a new kind of flow equation itself. More specifically,
consider the right hand side of the first order flow equation \eqref{eq:flowequationGammafirst}, in a slightly generalized form
\begin{equation}
  \text{B}_{k,l}[\Phi] = \frac{1}{2}\left(\frac{d}{dk^2} R_{kAB} \right) ((\Gamma_{k}^{(2)}[\Phi] + R_{k} + Q_l )^{-1})^{AB}.
\label{eq:def:BklFunctional}
\end{equation}
We have introduced here an additional regulator term $Q_l$ very similar in spirit to the standard regulator $R_k$, but now with a parameter $l$. It is assumed that $Q_l=0$ for $l=0$ and in that case one recovers the right hand side of the flow equation \eqref{eq:flowequationGammafirst},
\begin{equation}
  \frac{\partial}{\partial k^2} \Gamma_k[\Phi] = \text{B}_{k,0}[\Phi].
  \label{eq:flowGammaBkl}
\end{equation}

\subsection{Limit of very large $l$}
One can assume that the eigenvalues of $Q_l$ grow with $l$, typically like additional contributions to squared masses $\sim l^2$.

In the absence of ultraviolet divergences, or in the presence of an explicit UV regulator, we would for $l\to \infty$ have $Q_l \to \infty$ and therefore naively find
\begin{equation}
  \lim_{l\to \infty} \text{B}_{k,l}[\Phi] = 0.
  \label{eq:limBkl}
\end{equation}
However, we need to be very careful with this condition in the presence of ultraviolet divergences.

\subsection{Functional of modified inverse propagator}
Besides the regulator derivative $\frac{d}{dk^2} R_{kAB}$, which we assume to be independent of $\Phi$ and $l$, and to depend at most on $k$ in a simple and known way, we consider $B_{k,l}$ as a functional of the operator combination
\begin{equation}
   (\Gamma_k^{(2)}[\Phi]+R_k+Q_l)_{AB} = (\tilde\Gamma_k^{(2)}[\Phi] + Q_l)_{AB} = \frac{\delta^2}{\delta \Phi^A\delta \Phi^B} \tilde \Gamma_k[\Phi] + (Q_l)_{AB}.
\label{eq:assumptionsFunctionalFormOfB}
\end{equation}
This is the inverse propagator of the theory described by the Legendre transform $\tilde \Gamma_k[\Phi]$, to which a field-independent matrix or operator $(Q_l)_{AB}$ has been added.

The definition of the trace in eq.\ \eqref{eq:def:BklFunctional} is actually somewhat formal and heuristic, at least for mass-type regulators $R_k$ and $Q_l$, in the sense that there might be ultraviolet divergences. In the following we develop a method to circumvent the problem with these ultraviolet divergences, by first considering the flow of $\text{B}_{k,l}[\Phi]$ with $l$ and then integrating it. This will lead to a better definition of $\text{B}_{k,l}[\Phi]$, and therefore the flow equation for $\Gamma_k[\Phi]$. For that method we will need an assumption:
\begin{equation}
\begin{minipage}{0.8\linewidth}
$\text{B}_{k,l}[\Phi]$ \textit{ is a functional of the combination in eq.\ \eqref{eq:assumptionsFunctionalFormOfB}} \\ \textit{only, and not of individual terms entering the sum.}
\end{minipage}
\label{eq:assumption}
\end{equation}
We will see below how this assumption allows to find flow equations that are unique up to a small number of parameters.

\subsection{Secondary flow with new parameter $l$}
We now consider a flow equation for the dependence of $\text{B}_{k,l}[\Phi]$ on the parameter $l$,
\begin{equation}
  \frac{\partial}{\partial l^2}\text{B}_{k,l}[\Phi] = - \frac{1}{2} \left(\frac{d}{dk^2} R_{kAB} \right) \left(\frac{d}{dl^2} Q_{lCD} \right) ((\Gamma_{k}^{(2)}[\Phi] + R_{k} + Q_l )^{-1})^{AC} ((\Gamma_{k}^{(2)}[\Phi] + R_{k} + Q_l)^{-1})^{DB}.
\label{eq:flowBkl}
\end{equation}
Diagrammatically, this can be written as
\begin{equation}
  \frac{\partial}{\partial l^2} \text{B}_{k,l}[\Phi] = - \frac{1}{2} \;
  \begin{tikzpicture}[baseline=(base)]
    \node (base) at (0,-0.6ex) {};
    \begin{feynman}
      \vertex (a1);
      \vertex[above=0.4cm of a1] (b1);
      \vertex[right=0.4cm of a1] (c1);
      \vertex[below=0.4cm of a1] (d1);
      \vertex[left=0.4cm of a1] (e1);
      \diagram*{
        (b1) -- [quarter left] (c1),
        (c1) -- [quarter left] (d1),
        (d1) -- [quarter left] (e1),
        (e1) -- [quarter left] (b1),
        };
      \draw[fill=black] (b1) circle [radius=0.08cm];
      \draw[fill=gray] (d1) circle [radius=0.08cm];
    \end{feynman}
  \end{tikzpicture}
\end{equation}
where the black dot represents the standard regulator derivative $dR_k/dk^2$, and the gray dot the derivative $dQ_l/dl^2$.

By solving \eqref{eq:flowBkl} we can determine $B_{k,l}[\Phi]$, including the point $l=0$, which in turn determines the flow of the effective action according to \eqref{eq:flowGammaBkl}.

We also note that the flow equation \eqref{eq:flowBkl} resembles very closely one of the terms appearing in the second order flow equation \eqref{eq:flowequationGammaSecond}.

One could easily take further derivatives of \eqref{eq:flowBkl} with respect to $l^2$ which would lead to one-loop expressions with several propagators and regulator derivative insertions.

One advantage of the secondary flow, or ``flow of the flow equation'' in \eqref{eq:flowBkl} is that it has improved ultraviolet convergence properties compared to a direct application of the first order flow equation \eqref{eq:flowequationGammafirst}. The method allows to alleviate the need for ultraviolet regularization. We will demonstrate this advantage in more concrete situations below.

Of course, eventually one needs to integrate this secondary flow again, and additional integration constants will appear in that process. An important question is how these are to be determined, and what their physical significance is.

\section{Application to the effective potential}
\label{sec:ApplicationEffectivePotential}

To make the so far somewhat formal discussion more concrete, we will now discuss the flow of the effective potential in a scalar quantum field theory as an illustrating example.

\subsection{Standard first order flow equation in local potential approximation}
We specialize to a scalar field theory with global $O(N)$ symmetry in $d$ Euclidean dimensions. We study first an ansatz or truncation for the flowing action corresponding to a local potential approximation,
\begin{equation}
  \Gamma_{k}[\Phi] = \int d^d x \left\{ \frac{1}{2} \sum_n \delta^{\mu\nu} \partial_\mu \Phi_n(x) \partial_\nu \Phi_n(x) + U_{k}(\rho(x)) \right\},
  \label{eq:LocalPotentialApproximation}
\end{equation}
with the $O(N)$ invariant combination of fields
\begin{equation}
  \rho(x) = \frac{1}{2}\sum_{n=1}^N \Phi_n(x)^2.
\end{equation}
The flow of kinetic terms is neglected in the local potential approximation. The local potential approximation, which we employ here for illustration, has been previously studied in detail, see e.\ g.\ refs.\ \cite{Wetterich_1993_Average, Litim_1995_Approximate, Aoki_1996_Effectiveness, Berges_2002_Nonperturbative, Morris_2005_Equivalence, Bridle_2014_Local, Pelaez_2016_Ordered, Dupuis_2021_Nonperturbative}.

As infrared regulator we work with the choice $R_k = k^2$, which has the advantage that it preserves locality, unitarity, or Osterwalder-Schrader positivity in the Euclidean domain, as well as the $O(N)$ symmetry. On first sight we also face a disadvantage of this choice of regulator: it does not provide full ultraviolet regularization. In other words, some of the first order flow equations for $\Gamma_{k}[\Phi]$ obtained with this choice of regulator are not finite, and diverge when an additionally imposed UV regulator is removed.

We will discuss how different second order flow methods allow to address this problem.

In this section we first simply add an additional ultraviolet regularization by hand. This leads to the following flow equation for the effective potential $U_{k}(\rho)$,
\begin{equation}
\begin{split}
  \frac{\partial}{\partial k^2} U_{k}(\rho) = & \frac{1}{2(4\pi)^{d/2}} f_\Lambda^{(d)}\bigl(U_{k}^{\prime}(\rho) + 2 \rho U^{\prime\prime}_k(\rho)+k^2\bigr) + \frac{N-1}{2(4\pi)^{d/2}} f_\Lambda^{(d)}\bigl(U_{k}^{\prime}(\rho)+k^2\bigr).
\end{split}\label{eq:flowequationU}
\end{equation}
We have used here the abbreviation for loop functions
\begin{equation}
  f_\Lambda^{(d)}(z) = (4\pi)^{d/2} \int_\Lambda \frac{d^d p}{(2\pi)^d} \frac{1}{p^2+z}.
\label{eq:deffdLambdaz}
\end{equation}
For integer $d\geq 2$ this integral is not convergent at large momenta and we therefore need to impose some additional ultraviolet regularization at the scale $\Lambda$.

For example, when the UV regularization is done by the Schwinger proper time method one has
\begin{equation}
\begin{split}
  f_\Lambda^{(d)}(z) & = (4\pi)^{d/2} \int_\Lambda \frac{d^d p}{(2\pi)^d} \int_{1/\Lambda^2}^\infty d\tau \exp(-\tau [p^2+z]) \\
  & = \int_{1/\Lambda^2}^\infty d\tau \tau^{-d/2} \exp(-\tau z).
\end{split}\label{eq:flzDef}
\end{equation}
In the second line we have performed the momentum integration using Gaussian integration. The integral over $\tau$ leads, with a variable substitution $t=\tau z$, to an incomplete Gamma function,
\begin{equation}
  f_\Lambda^{(d)}(z) = z^{d/2-1}\Gamma(1-d/2, z/\Lambda^2).
\end{equation}
Definitions and some properties of incomplete Gamma function are collected in appendix \ref{app:GammaFunction}.

\subsubsection{Flow equation for large scale separation in four dimensions}

In the flow equation \eqref{eq:flowequationU} one may assume that the arguments obey
\begin{equation}
  U^{\prime}_k(\rho) + 2\rho U^{\prime\prime}_k(\rho) + k^2 \ll \Lambda^2, \quad\quad\quad U^{\prime}_k(\rho)  + k^2 \ll \Lambda^2.
\end{equation}
In this case one finds from the expansion of the incomplete Gamma function in eq.\ \eqref{eq:IncompleteGammam1xAsympt}, in $d=4$ dimensions
\begin{equation}
  f^{(4)}_\Lambda(z) \approx \Lambda^2 - (1-\gamma) z + z \ln(z/\Lambda^2)
= \Lambda^2 - z \ln\left(\frac{\Lambda^2 e^{1-\gamma}}{z}\right),
\end{equation}
and therefore
\begin{equation}
\begin{split}
  \frac{1}{k}\frac{\partial}{\partial k} U_k(\rho) =  & \frac{N \Lambda^2}{16\pi^2} - \frac{U^{\prime}_k(\rho) + 2\rho U^{\prime\prime}_k(\rho) + k^2}{16 \pi^2 } \ln\left( \frac{\Lambda^2 e^{1-\gamma}}{U_k^{\prime}(\rho)+ 2\rho U^{\prime\prime}_k(\rho) +k^2} \right) \\
  & - (N-1) \frac{ U^{\prime}_k(\rho) + k^2}{16 \pi^2} \ln\left(\frac{\Lambda^2e^{1-\gamma}}{U_k^{\prime}(\rho)+k^2} \right).
\end{split}\label{eq:flowUwithL}
\end{equation}
The field-independent term $\sim \Lambda^2$ on the right hand side can be dropped for most purposes. But there is still a logarithmic dependence on the UV regularization scale $\Lambda$, though. It is therefore not possible to take the limit $\Lambda \to \infty$ and to obtain a well defined first order flow equation for the effective potential in $d=4$ dimensions, with the Callan-Symanzik regulator and the local potential approximation.

The problem with this is that all terms that depend on $\Lambda$ also depend on the specific ultraviolet regularization scheme. The above flow equation is valid for the Schwinger proper time scheme, and large scale separation, but another regularization scheme would lead to other flow equations. In other words, the flow equations for finite UV regulator $\Lambda$ are not physically meaningful by themselves.

\subsection{Second order one-loop flow equation for the effective potential}
By evaluating eq. \eqref{eq:flowequationGamma2first} for constant field $\rho$ we obtain
\begin{equation}
\begin{split}
  \frac{\partial^2}{\partial (k^2)^2} U_k(\rho) = & -\frac{1}{2(4\pi)^{d/2}} \left[ 1+ \frac{\partial}{\partial k^2}  U^\prime_k(\rho) + 2 \rho \frac{\partial}{\partial k^2} U^{\prime\prime}_k(\rho) \right] g_\Lambda^{(d)}\bigl(U_{k}^{\prime}(\rho) + 2 \rho U^{\prime\prime}_k(\rho)+k^2\bigr) \\
  & - \frac{N-1}{2(4\pi)^{d/2}} \left[ 1+\frac{\partial}{\partial k^2}  U^\prime_k(\rho)\right] g_\Lambda^{(d)}\bigl(U_{k}^{\prime}(\rho)+k^2\bigr),
\end{split}\label{eq:flowequationUSecondFirstOrder}
\end{equation}
with
\begin{equation}
  g_\Lambda^{(d)}(z) = -\frac{\partial}{\partial z} f_\Lambda^{(d)}(z) = (4\pi)^{d/2} \int_\Lambda \frac{d^dp}{(2\pi)^d} \frac{1}{(p^2+z)^2}.
\label{eq:DefinitiongLambda}
\end{equation}
Equation \eqref{eq:flowequationUSecondFirstOrder} can also be obtained from a derivative of eq.\ \eqref{eq:flowequationU}. What we have obtained is a partial differential equation of maximally second order in $k$ (or $k^2$), but with first order derivatives being present, as well.

We note that $g_\Lambda^{(d)}(z)$ has improved convergence properties in the ultraviolet compared to $f_\Lambda^{(d)}(z)$. However, it still shows a logarithmic UV divergence. One finds for $d=4$ dimensions, in the Schwinger proper time regularization discussed above,
\begin{equation}
  g^{(4)}_\Lambda(z) = \Gamma(0, z/\Lambda^2).
\end{equation}
In the limit $z\ll \Lambda$ this becomes
\begin{equation}
  g^{(4)}_\Lambda(z) = \ln\left(\frac{\Lambda^2 e^{-\gamma}}{z}\right) = \ln(\Lambda^2 e^{-\gamma}) - \ln(z).
\end{equation}
Because this is multiplied with a field-dependent term in eq.\ \eqref{eq:flowequationUSecondFirstOrder}, one cannot simply remove the dependence on $\Lambda$ by subtracting a field-independent term from $U_k(\rho)$.

The issue can be traced back to the high-momentum limit of the scale derivative of the field-dependent two-point function, $\frac{\partial}{\partial k}\Gamma_k^{(2)}[\Phi](p)$, entering the last term on the right hand side of eq.\ \eqref{eq:flowequationGamma2first}. In a truncation where $\frac{\partial}{\partial k}\Gamma_k^{(2)}[\Phi](p)$ would vanish at large momenta, or at least become independent of $\Phi$, one could absorb all $\Lambda$-dependent terms in a $\rho$-independent and irrelevant constant part in $U_k(\rho)$. This would lead to a finite second-order flow equation!

In the context of the derivative expansion this means to set $g^{(4)}_\Lambda(z)  = - \ln(z)$ which results in the flow equation
\begin{equation}
\begin{split}
  \frac{\partial^2}{\partial (k^2)^2} U_k(\rho) = & \frac{1}{32 \pi^2} \left[ 1+ \frac{\partial}{\partial k^2}  U^\prime_k(\rho) + 2 \rho \frac{\partial}{\partial k^2} U^{\prime\prime}_k(\rho) \right] \ln\bigl(U_{k}^{\prime}(\rho) + 2 \rho U^{\prime\prime}_k(\rho)+k^2\bigr) \\
  & + \frac{N-1}{32 \pi^2} \left[ 1+\frac{\partial}{\partial k^2}  U^\prime_k(\rho)\right] \ln\bigl(U_{k}^{\prime}(\rho)+k^2\bigr).
\end{split}\label{eq:flowequationUSecondFirstOrderApprox1}
\end{equation}
This is now a completely finite flow equation. Because it is second-order, one needs to specify a larger set of initial or boundary conditions, for example in the form of $U_k(\rho)$ and $\frac{\partial}{\partial k^2} U_k(\rho)$.

\subsection{Second order two-loop flow equation for the effective potential}

Let us also explore the case where we take eq.\ \eqref{eq:flowequationGammaSecond} as a starting point and evaluate it for constant $\rho$. We need to evaluate two-loop expressions, which leads to
\begin{equation}
\begin{split}
  \frac{\partial^2}{\partial (k^2)^2} U_k(\rho) = & -\frac{1}{2(4\pi)^{d/2}} g_\Lambda^{(d)}\bigl(U_{k}^{\prime}(\rho) + 2 \rho U^{\prime\prime}_k(\rho)+k^2\bigr) - \frac{N-1}{2(4\pi)^{d/2}} g_\Lambda^{(d)}\bigl(U_{k}^{\prime}(\rho)+k^2\bigr) \\
  & + \frac{1}{4(4\pi)^d} \left[ \left( 3 U_k^{\prime\prime}(\rho) + 12 \rho U_k^{(3)}(\rho) + 4 \rho^2 U^{(4)}(\rho) \right) g_\Lambda^{(d)}\bigl( U^\prime_k(\rho) + 2\rho U^{\prime\prime}_k(\rho) + k^2 \bigr)^2 \right. \\
  & + 2(N-1)\left( U^{\prime\prime}_k(\rho) + 2 \rho U^{(3)}_k(\rho) \right) g_\Lambda^{(d)}\bigl( U^\prime_k(\rho) + 2\rho U^{\prime\prime}_k(\rho) + k^2 \bigr) g_\Lambda^{(d)}\bigl( U^\prime_k(\rho) + k^2 \bigr) \\
  & + \left. (N^2-1) U^{\prime\prime}_k(\rho) g_\Lambda^{(d)}\bigl( U^\prime_k(\rho) + k^2 \bigr)^2   \right] \\
  & - \frac{1}{(4\pi)^d} \rho \left( 3 U^{\prime\prime}_k(\rho) + 2 \rho U^{(3)}_k(\rho) \right)^2 \\
  & \times h_\Lambda^{(d)}\bigl( U^\prime_k(\rho) + 2 \rho U^{\prime\prime}_k(\rho) + k^2, U^\prime_k(\rho) + 2 \rho U^{\prime\prime}_k(\rho) + k^2, U^\prime_k(\rho) + 2 \rho U^{\prime\prime}_k(\rho) + k^2 \bigr) \\
  & - \frac{N-1}{(4\pi)^d}\rho \left(  U^{\prime\prime}_k(\rho) \right)^2 h_\Lambda^{(d)}\bigl( U^\prime_k(\rho) + 2 \rho U^{\prime\prime}_k(\rho) + k^2, U^\prime_k(\rho) + k^2, U^\prime_k(\rho) + k^2 \bigr) \\
  & - \frac{N-1}{(4\pi)^d}2 \rho \left(  U^{\prime\prime}_k(\rho) \right)^2 h_\Lambda^{(d)}\bigl( U^\prime_k(\rho) + k^2, U^\prime_k(\rho) + 2 \rho U^{\prime\prime}_k(\rho)  + k^2, U^\prime_k(\rho) + k^2 \bigr).
\end{split}\label{eq:flowequationUSecondOrderNoFirstOrder}
\end{equation}
Here we are using the two-loop function for the sunset diagram with two regulator insertions \cite{Ananthanarayan_2020_New},
\begin{equation}
\begin{split}
  h_\Lambda^{(d)}(x,y,z) = & (4\pi)^d \int_\Lambda \frac{d^d p d^d q}{(2\pi)^{2d}} \frac{1}{[(p+q)^2 + x] [p^2+y]^2 [q^2+z]^2} \\
  = & \int_{1/\Lambda^2}^\infty da \, db \, dc \frac{bc \, e^{-ax-by-cz}}{(ab+ac+bc)^{d/2}},
\end{split}
\end{equation}
using again a Schwinger proper time regularization in the second line. In fact, this diagram is UV finite in $d=4$ dimensions, so one could take $\Lambda\to \infty$ here. However, $g_\Lambda^{(d)}(x)$ as defined in eq.\ \eqref{eq:DefinitiongLambda} has a logarithmic UV divergence in $d=4$ dimensions. For the one-loop terms in the first line of eq.\ \eqref{eq:flowequationUSecondOrderNoFirstOrder} this is little problematic, because the UV divergence concerns only a $\rho$-independent term in the effective potential. For the two-loop ``figure of eight'' term in the second, third and fourth line this is different, however, and the logarithmic divergences are also appearing in field-dependent terms.

In other words, what we have obtained is a flow equation of second order where the right hand side is fully determined by the effective potential $U_k(\rho)$ and its derivatives with respect to $\rho$. The UV convergence properties are better than for the first order flow equation in \eqref{eq:flowequationU}, but logarithmic divergences in $d=4$ dimensions are still present.

We note here that eq.\ \eqref{eq:flowequationUSecondOrderNoFirstOrder} differs from an equation one could derive by taking a further derivative of the first order flow equation in \eqref{eq:flowequationU}, leading to eq.\ \eqref{eq:flowequationUSecondFirstOrder}, and then inserting further derivatives of eq.\ \eqref{eq:flowequationU} with respect to $\rho$ to replace the first scale derivatives on the right hand side. This is of course a consequence of the truncation. For an exact solution to the flow equation one could simply take another derivative.

This leads to an interesting idea: second order flow equations could be used for a systematic test of approximation schemes! For a quantitatively precise truncation, the second order two-loop flow equation \eqref{eq:flowequationGammaSecond} needs to be obeyed, as well\footnote{For alternative methods to test truncation schemes see \cite{Knorr_2021_Exact, Pawlowski_2017_Physics, DePolsi_2022_Regulator}.}.

\subsection{Secondary flow for the effective potential}

\subsubsection{Differential equations for the flow of the flow}
We now aim to apply the secondary flow method, as introduced in section~\ref{sec:secondaryFlowMethod}, to determine the right hand side of the flow equation for the effective potential. We choose $R_k$ and $Q_l$ to be $O(N)$ symmetric, $R_k = k^2$ and $Q_l = l^2$.

The first order flow equation is given by
\begin{equation}
  \frac{\partial}{\partial k^2} U_k(\rho) = b_{k,0}\bigl(U_k^\prime(\rho)+ 2 \rho U_k^{\prime\prime}(\rho), U_k^\prime(\rho) \bigr),
\end{equation}
with the function $b_{k,l}(x,y)$ defined, in the presence of an ultraviolet regularization scheme at the scale $\Lambda$,
\begin{equation}
\begin{split}
   b_{k,l}(x,y)
   = & \frac{1}{2} \frac{1}{(4\pi)^{d/2}} f_\Lambda^{(d)}\bigl(x+k^2+l^2\bigr) + \frac{1}{2} \frac{N-1}{(4\pi)^{d/2}} f_\Lambda^{(d)}\bigl(y+k^2+l^2\bigr).
\end{split}\label{eq:flowequationUSecondary}
\end{equation}
So far, this still depends on the specific choice of ultraviolet regularization scheme as indicated by the dependence on $\Lambda$.

The idea is now to take derivatives with respect to $l$ (or $l^2$) until one arrives at a function for which the limit $\Lambda\to\infty$ can be taken, and to then integrate the result. In practice this means to characterize $f^{(d)}(z)$ through differential equations, and then to integrate these with appropriate boundary conditions. The differential equations are then independent of ultraviolet regularization scheme, but the final expressions obtained for $b_{k,l}(x,y)$ will have a dependence on additional integration constants.

We assume that one can write in this scheme
\begin{equation}
  b_{k,l}(x,y) = \frac{1}{2} \frac{1}{(4\pi)^{d/2}} f^{(d)}\bigl(x+k^2+l^2\bigr) + \frac{1}{2} \frac{N-1}{(4\pi)^{d/2}} f^{(d)}\bigl(y+k^2+l^2\bigr),
\end{equation}
and aim to determine $f^{(d)}(z)$.\footnote{In principle, $b_{k,l}(x,y)$ could depend on $k$, $l$, $x$ and $y$ in a more complicated way. However, from the assumption \eqref{eq:assumption} it follows that this is not the case.}

From the definition in eq.\ \eqref{eq:deffdLambdaz} we find, still in the presence of a finite UV regulator scale $\Lambda$,
\begin{equation}
  \frac{\partial^n}{\partial z^n}f_\Lambda^{(d)}(z) = (-1)^n (4\pi)^{d/2} \int_\Lambda \frac{d^d p}{(2\pi)^d} \frac{n!}{(p^2+z)^{n+1}}.
\end{equation}
Note that this is convergent in the UV for $n+1>d/2$, so once enough derivatives have been taken. One finds then, once the limit $\Lambda \to \infty$ has been taken,
\begin{equation}
  \frac{\partial^n}{\partial z^n}f^{(d)}(z) = (-1)^n \frac{\Gamma(1+n-d/2)}{z^{1+n-d/2}}.
\label{eq:flzDerIntPerformed}
\end{equation}
This is finally independent of the ultraviolet regularization scheme! Observe that for $n+1-d/2>0$ the right hand side vanishes for $z\to \infty$. One may then integrate from that point, which of course introduces corresponding integration constants. This is best discussed with concrete values for $d$.

\subsubsection{Solution for different numbers of spacetime dimensions}
For example, in the case $d=2$ one integration, starting from \eqref{eq:flzDerIntPerformed} with $n=1$,
\begin{equation}
  \frac{\partial}{\partial z} f^{(2)}(z) = -\frac{1}{z},
\end{equation}
leads to
\begin{equation}
  f^{(2)}(z) = \ln\left( \frac{L_1^2}{z} \right),
\end{equation}
where $L_1$ is an integration constant of dimension mass.

Similarly, for $d=3$ one integration from \eqref{eq:flzDerIntPerformed} with $n=1$,
\begin{equation}
  \frac{\partial}{\partial z} f^{(3)}(z) = - \frac{\sqrt{\pi}}{\sqrt{z}}
\end{equation}
leads to
\begin{equation}
  f^{(3)}(z) = 2\sqrt{\pi}\left[L_1 -\sqrt{z}\right],
\end{equation}
where again $L_1$ is an integration constant of dimension mass.

For $d=4$ one must start with \eqref{eq:flzDerIntPerformed} with $n=2$,
\begin{equation}
  \frac{\partial^2}{\partial z^2} f^{(4)}(z) = \frac{1}{z},
\end{equation}
and integrate twice, leading to
\begin{equation}
  f^{(4)}(z) = L_2^2- z \left[1+\ln\left( \frac{L_1^2}{z} \right)\right],
\end{equation}
where we have now two integration constants $L_1$ and $L_2$, each with dimensions of mass. Note that this is consistent with \eqref{eq:flowUwithL}, albeit somewhat more general.

The resulting flow equation for the effective potential is
\begin{equation}
\begin{split}
  \frac{\partial}{\partial k^2} U_k(\rho) =  & N \frac{ L_2^2}{32\pi^2} - \frac{U^{\prime}_k(\rho) + 2\rho U^{\prime\prime}_k(\rho) + k^2}{32 \pi^2 }  \left[1+\ln\left( \frac{L_1^2}{U_k^{\prime}(\rho)+ 2\rho U^{\prime\prime}_k(\rho) +k^2} \right) \right] \\
  & - (N-1) \frac{ U^{\prime}_k(\rho) + k^2}{32 \pi^2} \left[1 + \ln\left(\frac{L_1^2}{U_k^{\prime}(\rho)+k^2} \right) \right] .
\end{split}\label{eq:flowUwithLambda12}
\end{equation}
Note that $L_2$ affects only a field-independent part of $U_k(\rho)$, while $L_1$ affects in general all powers in $\rho$.

\subsubsection{Physical significance of the scheme and integration constants}
The scheme we have developed here has a compelling physical interpretation:
\begin{enumerate}
	\item With the infrared regulator $R_k=k^2$, the first order flow equation for the flowing action \eqref{eq:flowequationGammafirst}, or the analog in eq.\ \eqref{eq:flowequationU} for the effective potential, describes essentially the change in the quantum effective action when the mass-squared parameter of the microscopic model is varied.
	\item The microscopic model is defined at some ultraviolet scale $\Lambda$ but the precise ``physical'' implementation of the regularization is not known.
	\item It is important that the flowing action $\Gamma_k[\Phi]$ as well as its flow $\frac{\partial}{\partial k^2} \Gamma_k[\Phi]$, or their analogs $U_k(\phi)$ and $\frac{\partial}{\partial k^2}U_k(\phi)$ are also completely physical in this sense, albeit in general dependent on the regularization at the scale $\Lambda$.
	\item Adding an additional mass-squared term $l^2$ to the inverse propagator amounts to studying a class of \textit{quantum effective actions} that differ only in a mass-squared term, $\Delta\Gamma_k[\Phi] = \frac{l^2}{2} \int d^d x \Phi(x)^2 $, while the rest of the flowing action agrees. It is an assumption that such a class of models exists (we have no explicit microscopic construction for them), but this is very plausible.
	\item Also the derivatives of the flow $\frac{\partial}{\partial k} \Gamma_k[\Phi]$ with respect to $l$ or $l^2$ are fully physical in this sense. As argued before, these derivatives depend weakly enough on $\Lambda$, so that one can take the limit $\Lambda\to \infty$.
	\item Integrating the flow with $l$ again leads back to the original flow equation $\frac{\partial}{\partial k} \Gamma_k[\Phi]$. The integration constants that need to be introduced in that process parametrize the ignorance about the precise definition of the model at the microscopic scale.
\end{enumerate}
Interestingly, the knowledge of the flowing action $\Gamma_k[\Phi]$ itself is not enough to fix all the necessary parameters governing the flow $\frac{\partial}{\partial k} \Gamma_k[\Phi]$! Information about ultraviolet physics enters not only through the coupling constants in $\Gamma_k[\Phi]$ but also in the form of integration constants $L_1$ and $L_2$.

The integration constants $L_1$ and $L_2$ are dominated by fluctuations in the ultraviolet. This is the regime where an effective field theory description cannot be trusted, and $L_1$ and $L_2$ are therefore not predictable within the infrared theory itself. This is similar as for relevant or marginal coupling constants, which need to be fixed by experimental measurements in the infrared regime.

In principle, if one could vary the mass-squared parameter of the microscopic model, and could observe the change in the parameters of the flowing action, one could access the additional parameters $L_1$ and $L_2$. In practice this seems impossible, except in numerical experiments.

\subsection{Taylor expansion and flow of couplings}
To simplify things, let us study a Taylor expansion of the effective potential around a minimum $\rho_0$,
\begin{equation}
  U_{k}(\rho) = u_k + m_k^2 [\rho-\rho_0] + \frac{\lambda_k}{2} [\rho-\rho_0]^2 + \frac{\gamma_k}{3!} [\rho-\rho_0]^3 + \frac{\zeta_k}{4!} [\rho-\rho_0]^4+ \frac{\xi_k}{5!} [\rho-\rho_0]^5+\ldots,
  \label{eq:TaylorExpansionFlowingPotential}
\end{equation}
For $\rho_0>0$ one has $m_k^2=0$ while $m_k^2>0$ is allowed for $\rho_0=0$. We concentrate on the latter case for simplicity.


By inserting the expansion \eqref{eq:TaylorExpansionFlowingPotential} in \eqref{eq:flowUwithLambda12} and expanding we find in $d=4$ dimensions
\begin{equation}
\begin{split}
  \frac{d}{dk^2} u_k = & \frac{N}{32\pi^2} \left[ L_2^2 - (m_k^2+k^2) \left[1+\ln\left( \frac{L_1^2}{m_k^2+k^2} \right)  \right] \right], \\
  \frac{d}{dk^2} (m_k^2 + k^2) = & 1 - \frac{[N+2]\lambda_k}{32\pi^2} \ln\left( \frac{L_1^2}{m_k^2+k^2} \right), \\
  \frac{d}{dk^2} \lambda_k = & \frac{[N+8]\lambda_k^2}{32\pi^2[m_k^2+k^2]} - \frac{[N+4] \gamma_k}{32\pi^2} \ln\left( \frac{L_1^2}{m_k^2+k^2} \right), \\
  \frac{d}{dk^2} \gamma_k = &
  -\frac{[N+26]\lambda_k^3}{32\pi^2[m_k^2+k^2]^2}  + \frac{[3N+42]\lambda_k \gamma_k}{32\pi^2 [m_k^2+k^2]} -  \frac{[N+6]\zeta_k}{32\pi^2} \ln\left( \frac{L_1^2}{m_k^2+k^2} \right).
\end{split}\label{eq:FlowEqSet2}
\end{equation}
These equations are solved (for given $\zeta_k$, e.g. $\zeta_k=0$) as usual, with some boundary conditions for $u_k$, $m_k^2$,  $\lambda_k$, and $\gamma_k$. The usual procedure is to fix $m_k^2$ and $\lambda_k$, and also $u_k$ by renormalization conditions at $k=0$ and $\gamma_k$ at very large $k$ to a small or vanishing value. Because the flow of $\gamma_k$ and similar irrelevant couplings is infrared dominated, its value can be predicted at $k=0$ \cite{Berges_2002_Nonperturbative}.

But how should one fix $L_1$ and $L_2$ ? For $L_2$ the question is not very pressing because it only affects the flow equation for $u_k$ which does not enter anywhere else. However, $L_1$ appears in all flow equations!

\subsubsection{Perturbative solution}
To connect to intuition from perturbation theory, we discuss here the solution to flow equations \eqref{eq:FlowEqSet2} to leading order in a perturbative expansion in $\lambda_k$. One expects in the vicinity of a Gaussian fixed point $\gamma_k \sim \lambda_k^3$ and $\zeta_k \sim \lambda_k^4$. To lowest order, the flow of the mass can be neglected, $m_k^2=m^2 = m_0^2$. The flow equation for the quartic coupling,
\begin{equation}
  \frac{d}{dk^2}\lambda_k = \frac{[N+8]\lambda_k^2}{32\pi^2 [m_k^2 + k^2]},
\end{equation}
can be directly integrated, with the solution
\begin{equation}
  \lambda(k) = \frac{1}{\frac{1}{\lambda(\Lambda)} + \frac{N+8}{32\pi^2} \ln\left( \frac{m_0^2+\Lambda^2}{m_0^2+k^2} \right)}.
\end{equation}
One observes the typical Landau pole behavior leading to an upper bound on $\lambda(0)$ for given initialization scale $\Lambda$, under the assumption that $\lambda(k)\geq 0$ at all scales $k$.

With this lowest order solution, the flow equation for $m_k^2+k^2$ becomes
\begin{equation}
  \frac{d}{dk^2} (m_k^2+k^2) = 1 - \frac{[N+2] \lambda_k}{32\pi^2} \ln\left( \frac{L_1^2}{m_k^2 + k^2} \right) = 1 - \frac{[N+2]}{[N+8]} \frac{\ln\left(\frac{L_1^2}{m_k^2 + k^2}\right)}{\frac{32\pi^2}{[N+8]\lambda(\Lambda)}+\ln\left(\frac{m_0^2 + \Lambda^2}{m_0^2 + k^2}\right)}.
\end{equation}
The ratio on the right is small for small coupling $\lambda(\Lambda)$, but could become of order unity when the coupling becomes large.

One can also first neglect the flow of $\lambda_k$ and set $\lambda_k = \lambda_0$ in the flow of $m_k^2$. That yields
\begin{equation}
	\frac{d}{dk^2} (m_k^2+k^2) = 1 - \frac{[N+2]\lambda_0}{32 \pi^2} \ln\left( \frac{L_1^2}{m_k^2 + k^2} \right).
\end{equation}
A perturbative solution for small $k^2$ is of the form
\begin{equation}
	m_k^2 + k^2 = m_0^2 + k^2 \left[ 1-  \frac{[N+2]\lambda_0}{32 \pi^2} \ln\left( \frac{L_1^2}{m_0^2} \right) \right] + \ldots.
\end{equation}
This reveals an interesting choice for $L_1^2$: For $L_1^2=m_0^2$ the flow of $m_k^2$ vanishes at $k=0$ and up to terms of order $k^4$ one has the effective mass $m_k^2+k^2 = m_0^2 + k^2$.

\subsubsection{Flow with mass}
In section \ref{sec:RelationCallanToFunctionalFormulation} we discussed the relation between Callan's renormalization scheme and the functional formulation. In that context we considered a flow with the effective mass. It is instructive to discuss this formulation here in a more concrete way.

By combining the second and third line from \eqref{eq:FlowEqSet2} one finds with $m=\sqrt{m_k^2+k^2}$,
\begin{equation}
\begin{split}
	\beta_\lambda = & m \frac{d\lambda}{d m} = 2(m_k^2+k^2)\frac{1}{\frac{d(m_k^2+k^2)}{dk^2}} \frac{d\lambda_k}{dk^2} \\
	= & \frac{\frac{[N+8]\lambda_k^2}{16\pi^2} - \frac{[N+4] \gamma_k}{16\pi^2} (m_k^2+k^2) \ln\left( \frac{L_1^2}{m_k^2+k^2} \right)}{1 - \frac{[N+2]\lambda_k}{32\pi^2} \ln\left( \frac{L_1^2}{m_k^2+k^2} \right)}.
\end{split}
\end{equation}
Note that this flow equation is independent of how the flow trajectory is parametrized, i.\ e.\ redefinitions of the parameter $k$.

The first term in the numerator in the last line is the leading expression in a perturbative expansion for small $\lambda_k$. The dependence on the integration constant $L_1$ appears in higher orders, but seems unavoidable, at least in the local potential approximation.

\section{Discussion, conclusions and outlook}
\label{sec:discussion}

To summarize, we have discussed here formulations of functional renormalization group equations of higher orders, beyond the first order one-loop flow equations mainly used in the field so far.

The primary motivation was to improve the ultraviolet convergence properties, similar to what happens in Callan's perturbative renormalization scheme with higher-order $\theta$-insertions. This would help to formulate flow equations with an infrared regulator that has the simple form of an additional contribution to squared masses. This, in turn, would lead to a flow that preserves symmetries, locality and most importantly unitarity of the theory (or Osterwalder-Schrader reflection positivity in the Euclidean domain).

Our first approach was to take further derivatives of the flowing action $\Gamma_k[\Phi]$ with respect to the flow parameter $k$. That led to very interesting new exact flow equations. For example, at second order, one finds two possible formulations. One has a one-loop structure, but with first derivatives of the inverse propagator with respect to the flow parameter appearing on the right hand side. The second formulation uses the standard first order flow equation for these terms and leads then to a second-order flow equation with one-loop and two-loop structures on the right hand side.

These equations are very interesting by themselves. Being additional exact functional renormalization group equations, they can be used as a starting point for approximate calculations with a focus on characteristic two-loop effects. Another possible use is for checking truncation schemes. The new higher-order flow equations need to be obeyed by the full flowing action, but for approximate solutions one can find deviations. For quantitatively precise truncation schemes these deviations should be minimal.

What the higher-order flow equations could not do is to improve the ultraviolet convergence properties enough to make flow equations finite, for example for a scalar field theory in $d=4$ dimensions. There are always subdiagrams with tadpole-structure with remaining logarithmic ultraviolet divergences.

This is in fact consistent with the expectations from Callan's perturbative scheme \cite{Callan_1993_Introduction, Mooij_2023_QFT, Mooij_2023_Finite, Mooij_2024_Effective} and a recent extension to non-renormalizable theories \cite{Ageeva_2025_Nonrenormalizable}. In the realm of perturbation theory one needs only a finite number of $\theta$-insertions, or derivatives with respect to the microscopic squared mass, to obtain finite expressions in perturbation theory (two for $\phi^4$ scalar field theory in $d=4$ dimensions). However, for non-renormalizable theories this changes already. Here one would need infinitely many such $\theta$ insertions if one would aim to do calculations to all orders in perturbation theory. Only for a calculation to a finite order in perturbation theory does the number of necessary $\theta$-insertions become finite again.

Non-perturbative functional renormalization group calculations are similar to non-renormalizable theories in that regard. One makes an ansatz for the effective action that contains not only the relevant and marginal terms in the vicinity of a Gaussian fixed point, but also technically irrelevant operators. As a consequence, the higher-order exact flow equations are not becoming ultraviolet finite. Additional truncation steps would be necessary to reach that goal, for example an additional expansion to some order in $\hbar$ could do the job.

In order to solve the ultraviolet convergence issue, we have also discussed another scheme: the secondary flow. The idea was here to take further derivatives of the standard first-order one-loop renormalization group flow equation with respect to effective squared-mass parameters. This leads to secondary flow equations, or a flow of the flow equation, with an additional parameter $l$. When enough derivatives have been taken, these secondary flow equations become ultraviolet convergent and finite, and can be evaluated without additional ultraviolet regularization.

In order to come back to the standard first order flow equation one must integrate the secondary flow. This leads to the appearance of integration constants. In a concrete application for the flow of the effective potential of an $O(N)$-model in $d=4$ dimensions we found that two integration constants $L_1$ and $L_2$ with dimensions of mass were needed. These should be seen as parameters that are in principle fixed by the microscopic or ultraviolet properties of a quantum field theory, very much like relevant coupling constants. It is a bit surprising that the knowledge of the flowing action $\Gamma_k[\Phi]$ itself is not enough to fix all parameters of this kind -- but that's just how it is. A systematic study of the scheme dependence encoded in $L_1$ and $L_2$, and its relation to other perturbative and non-perturbative  formulations of the renormalization group (e.\ g.\ \cite{Baldazzi_2021_Functional, Beretta_2026_Functional, Alexander_2019_Alternative}) would be valuable. In that context one should also investigate whether integration constants like $L_1$ and $L_2$ can flow with the renormalization group, as well.

The secondary flow method as such looks promising, and one can expect that it can be well applied to many different physics situations and models. Besides scalar field theory it would be particularly interesting to study fermionic theories and gauge theories \cite{Gies_2012_Introduction, Metzner_2012_Functional} in a next step. The Callan-Symanzik-type regulator could show particular advantages for the study of real-time phenomena, bound states, and analytic continuation \cite{Ellwanger_1993_Flow, Gies_2002_Renormalization-a, Floerchinger_2012_Analytic, Kamikado_2014_Real-time, Jakovac_2019_Bound, Fehre_2023_Lorentzian, Braun_2023_Renormalised, Horak_2024_Scalar, DAngelo_2024_Algebraic} of functional renormalization group equations from the Euclidean domain to real frequencies and times. Also the application to curved spacetimes \cite{Baume_2014_Local, Guilleux_2015_Quantum, Markkanen_2018_1-loop} and the renormalization of interacting quantum field theories in non-trivial geometries would be a very interesting future direction.

\appendix

\section{Second-order flow equations for higher correlation functions}
\label{app:HigherDerivatives}

In the main text we presented the derivation of the second-order, two-loop  flow equation \eqref{eq:flowequationGammaSecond}. Functional derivatives of that equation with respect to $\Phi$ yield second-order flow equations for correlation functions, including the two-loop contributions. Unfortunately, but expectedly, these still contain subdivergences in the limit $\Lambda\to \infty$ as severe as those appearing in the first order flow equation.

In diagrammatic form one finds the second-order flow equations for the one-point function
\begin{equation}
\frac{\partial^2}{\partial k^2}  (\Gamma^{(1)}_{k}[\Phi])_A = - \frac{1}{2} \;
\begin{tikzpicture}[baseline=(base)]
  \node (base) at (0,-0.6ex) {};
  \begin{feynman}
    \vertex (a1);
    \vertex[above=0.4cm of a1] (b1);
    \vertex[right=0.4cm of a1] (c1);
    \vertex[below=0.4cm of a1] (d1);
    \vertex[left=0.4cm of a1] (e1);
    \vertex[below=0.2cm of d1] (extA) {\(A\)};
    \diagram*{
      (b1) -- [quarter left] (c1),
      (c1) -- [quarter left] (d1),
      (d1) -- [quarter left] (e1),
      (e1) -- [quarter left] (b1),
      (d1) -- (extA),
      };
      \draw[fill=white] (b1) circle [radius=0.08cm];
  \end{feynman}
\end{tikzpicture} \;
+ \;
\begin{tikzpicture}[baseline=(base)]
  \node (base) at (0,-0.6ex) {};
  \begin{feynman}
    \vertex (a1);
    \vertex[above=0.4cm of a1] (b1);
    \vertex[right=0.4cm of a1] (c1);
    \vertex[below=0.4cm of a1] (d1);
    \vertex[left=0.4cm of a1] (e1);
    \vertex[below=0.2cm of d1] (extA) {\(A\)};
    \diagram*{
      (b1) -- [quarter left] (c1),
      (c1) -- [quarter left] (d1),
      (d1) -- [quarter left] (e1),
      (e1) -- [quarter left] (b1),
      (d1) -- (extA),
      };
    \draw[fill=black] (c1) circle [radius=0.08cm];
    \draw[fill=black] (e1) circle [radius=0.08cm];
  \end{feynman}
\end{tikzpicture} \;
- \;
\begin{tikzpicture}[baseline=(base)]
  \node (base) at (0,-0.6ex) {};
  \begin{feynman}
    \vertex (a1);
    \vertex[above=0.2cm of a1] (f1);
    \vertex[below=0.2cm of a1] (g1);
    \vertex[right=0.2cm of f1] (f2);
    \vertex[left=0.2cm of f1] (f3);
    \vertex[right=0.2cm of g1] (g2);
    \vertex[left=0.2cm of g1] (g3);
    \vertex[right=0.4cm of a1] (c1);
    \vertex[left=0.4cm of a1] (e1);
    \vertex[below=0.2cm of g3] (extA) {\(A\)};
    \diagram*{
      (c1) -- [quarter right] (f2),
      (f2) -- [quarter right] (a1),
      (a1) -- [quarter right] (g2),
      (g2) -- [quarter right] (c1),
      (a1) -- [quarter right] (f3),
      (f3) -- [quarter right] (e1),
      (e1) -- [quarter right] (g3),
      (g3) -- [quarter right] (a1),
      (g3) -- (extA),
      };
    \draw[fill=black] (c1) circle [radius=0.08cm];
    \draw[fill=black] (e1) circle [radius=0.08cm];
  \end{feynman}
\end{tikzpicture} \;
+ \frac{1}{4} \;
\begin{tikzpicture}[baseline=(base)]
  \node (base) at (0,-0.6ex) {};
  \begin{feynman}
    \vertex (a1);
    \vertex[above=0.2cm of a1] (f1);
    \vertex[below=0.2cm of a1] (g1);
    \vertex[right=0.2cm of f1] (f2);
    \vertex[left=0.2cm of f1] (f3);
    \vertex[right=0.2cm of g1] (g2);
    \vertex[left=0.2cm of g1] (g3);
    \vertex[right=0.4cm of a1] (c1);
    \vertex[left=0.4cm of a1] (e1);
    \vertex[below=0.3cm of a1] (extA) {\(A\)};
    \diagram*{
      (c1) -- [quarter right] (f2),
      (f2) -- [quarter right] (a1),
      (a1) -- [quarter right] (g2),
      (g2) -- [quarter right] (c1),
      (a1) -- [quarter right] (f3),
      (f3) -- [quarter right] (e1),
      (e1) -- [quarter right] (g3),
      (g3) -- [quarter right] (a1),
      (a1) -- (extA),
      };
    \draw[fill=black] (c1) circle [radius=0.08cm];
    \draw[fill=black] (e1) circle [radius=0.08cm];
  \end{feynman}
\end{tikzpicture} \;
+ 2 \;
\begin{tikzpicture}[baseline=(base)]
  \node (base) at (0,-0.6ex) {};
  \begin{feynman}
    \vertex (a1);
    \vertex[above=0.4cm of a1] (b1);
    \vertex[right=0.4cm of a1] (c1);
    \vertex[below=0.4cm of a1] (d1);
    \vertex[left=0.4cm of a1] (e1);
    \vertex (p1) at (140:0.4cm);
    \vertex (p2) at (-35:0.4cm);
    \vertex (p3) at (230:0.4cm);
    \vertex (p4) at (55:0.4cm);
    \vertex[below=0.2cm of d1] (extA) {\(A\)};
    \diagram*{
      (b1) -- [quarter left] (c1),
      (c1) -- [quarter left] (d1),
      (d1) -- [quarter left] (e1),
      (e1) -- [quarter left] (b1),
      (p3) -- (p4),
      (d1) -- (extA),
      };
    \draw[fill=black] (p1) circle [radius=0.08cm];
    \draw[fill=black] (p2) circle [radius=0.08cm];
  \end{feynman}
\end{tikzpicture} \;
+ \frac{1}{2} \;
\begin{tikzpicture}[baseline=(base)]
  \node (base) at (0,-0.6ex) {};
  \begin{feynman}
    \vertex (a1);
    \vertex[above=0.4cm of a1] (b1);
    \vertex[right=0.4cm of a1] (c1);
    \vertex[below=0.4cm of a1] (d1);
    \vertex[left=0.4cm of a1] (e1);
    \vertex[below=0.2cm of d1] (extA) {\(A\)};
    \diagram*{
      (b1) -- [quarter left] (c1),
      (c1) -- [quarter left] (d1),
      (d1) -- [quarter left] (e1),
      (e1) -- [quarter left] (b1),
      (e1) -- (c1),
      (d1) -- (extA),
      };
    \draw[fill=black] (b1) circle [radius=0.08cm];
    \draw[fill=black] (a1) circle [radius=0.08cm];
  \end{feynman}
\end{tikzpicture} \;
- \;
\begin{tikzpicture}[baseline=(base)]
  \node (base) at (0,-0.6ex) {};
  \begin{feynman}
    \vertex (a1);
    \vertex[above=0.4cm of a1] (b1);
    \vertex[right=0.4cm of a1] (c1);
    \vertex[below=0.4cm of a1] (d1);
    \vertex[left=0.4cm of a1] (e1);
    \vertex[below=0.2cm of d1] (extA) {\(A\)};
    \diagram*{
      (b1) -- [quarter left] (c1),
      (c1) -- [quarter left] (d1),
      (d1) -- [quarter left] (e1),
      (e1) -- [quarter left] (b1),
      (b1) -- (d1),
      (d1) -- (extA),
      };
    \draw[fill=black] (c1) circle [radius=0.08cm];
    \draw[fill=black] (e1) circle [radius=0.08cm];
  \end{feynman}
\end{tikzpicture}
\end{equation}
Note that these would all vanish in a $Z_2$ symmetric situation where vertices $\Gamma^{(n)}$ with $n$ odd are absent.

For the two-point function we find diagrammatically
\begin{equation}
\frac{\partial^2}{\partial k^2}  (\Gamma^{(2)}_{k}[\Phi])_{AB} =
  - \frac{1}{2} \; 
  \begin{tikzpicture}[baseline=(base)]
    \node (base) at (0,-0.6ex) {};
    \begin{feynman}
      \vertex (a1);
      \vertex[above=0.4cm of a1] (b1);
      \vertex[right=0.4cm of a1] (c1);
      \vertex[below=0.4cm of a1] (d1);
      \vertex[left=0.4cm of a1] (e1);
      \vertex[below left=0.25cm and 0.15cm of d1] (extA) {\(A\)};
      \vertex[below right=0.25cm and 0.15cm of d1] (extB) {\(B\)};
      \diagram*{
        (b1) -- [quarter left] (c1),
        (c1) -- [quarter left] (d1),
        (d1) -- [quarter left] (e1),
        (e1) -- [quarter left] (b1),
        (d1) -- (extA),
        (d1) -- (extB),
        };
        \draw[fill=white] (b1) circle [radius=0.08cm];
    \end{feynman}
  \end{tikzpicture} \;
  + \; 
  \begin{tikzpicture}[baseline=(base)]
    \node (base) at (0,-0.6ex) {};
    \begin{feynman}
      \vertex (a1);
      \vertex[above=0.4cm of a1] (b1);
      \vertex[right=0.4cm of a1] (c1);
      \vertex[below=0.4cm of a1] (d1);
      \vertex[left=0.4cm of a1] (e1);
      \vertex[below left=0.25cm and 0.15cm of d1] (extA) {\(A\)};
      \vertex[below right=0.25cm and 0.15cm of d1] (extB) {\(B\)};
      \diagram*{
        (b1) -- [quarter left] (c1),
        (c1) -- [quarter left] (d1),
        (d1) -- [quarter left] (e1),
        (e1) -- [quarter left] (b1),
        (d1) -- (extA),
        (d1) -- (extB),
        };
      \draw[fill=black] (c1) circle [radius=0.08cm];
      \draw[fill=black] (e1) circle [radius=0.08cm];
    \end{feynman}
  \end{tikzpicture} \;
  - \; 
  \begin{tikzpicture}[baseline=(base)]
    \node (base) at (0,-0.6ex) {};
    \begin{feynman}
      \vertex (a1);
      \vertex[above=0.2cm of a1] (f1);
      \vertex[below=0.2cm of a1] (g1);
      \vertex[right=0.2cm of f1] (f2);
      \vertex[left=0.2cm of f1] (f3);
      \vertex[right=0.2cm of g1] (g2);
      \vertex[left=0.2cm of g1] (g3);
      \vertex[right=0.4cm of a1] (c1);
      \vertex[left=0.4cm of a1] (e1);
      \vertex[below left=0.25cm and 0.15cm of g3] (extA) {\(A\)};
      \vertex[below right=0.25cm and 0.15cm of g3] (extB) {\(B\)};
      \diagram*{
        (c1) -- [quarter right] (f2),
        (f2) -- [quarter right] (a1),
        (a1) -- [quarter right] (g2),
        (g2) -- [quarter right] (c1),
        (a1) -- [quarter right] (f3),
        (f3) -- [quarter right] (e1),
        (e1) -- [quarter right] (g3),
        (g3) -- [quarter right] (a1),
        (g3) -- (extA),
        (g3) -- (extB),
        };
      \draw[fill=black] (c1) circle [radius=0.08cm];
      \draw[fill=black] (e1) circle [radius=0.08cm];
    \end{feynman}
  \end{tikzpicture} \;
  - \; 
  \begin{tikzpicture}[baseline=(base)]
    \node (base) at (0,-0.6ex) {};
    \begin{feynman}
      \vertex (a1);
      \vertex[above=0.4cm of a1] (b1);
      \vertex[right=0.4cm of a1] (c1);
      \vertex[below=0.4cm of a1] (d1);
      \vertex[left=0.4cm of a1] (e1);
      \vertex[above=0.6cm of a1] (extB) {\(B\)};
      \vertex[below=0.6cm of a1] (extA) {\(A\)};
      \diagram*{
        (b1) -- [quarter left] (c1),
        (c1) -- [quarter left] (d1),
        (d1) -- [quarter left] (e1),
        (e1) -- [quarter left] (b1),
        (extB) -- (extA),
        };
      \draw[fill=black] (c1) circle [radius=0.08cm];
      \draw[fill=black] (e1) circle [radius=0.08cm];
    \end{feynman}
  \end{tikzpicture} \;
  + \frac{1}{4} \; 
  \begin{tikzpicture}[baseline=(base)]
    \node (base) at (0,-0.6ex) {};
    \begin{feynman}
      \vertex (a1);
      \vertex[above=0.2cm of a1] (f1);
      \vertex[below=0.2cm of a1] (g1);
      \vertex[right=0.2cm of f1] (f2);
      \vertex[left=0.2cm of f1] (f3);
      \vertex[right=0.2cm of g1] (g2);
      \vertex[left=0.2cm of g1] (g3);
      \vertex[right=0.4cm of a1] (c1);
      \vertex[left=0.4cm of a1] (e1);
      \vertex[above=0.3cm of a1] (extB) {\(B\)};
      \vertex[below=0.3cm of a1] (extA) {\(A\)};
      \diagram*{
        (c1) -- [quarter right] (f2),
        (f2) -- [quarter right] (a1),
        (a1) -- [quarter right] (g2),
        (g2) -- [quarter right] (c1),
        (a1) -- [quarter right] (f3),
        (f3) -- [quarter right] (e1),
        (e1) -- [quarter right] (g3),
        (g3) -- [quarter right] (a1),
        (extB) -- (extA),
        };
      \draw[fill=black] (c1) circle [radius=0.08cm];
      \draw[fill=black] (e1) circle [radius=0.08cm];
    \end{feynman}
  \end{tikzpicture}
\end{equation}
For simplicity we have concentrated here on those terms that contain only $Z_2$ symmetric vertices. Unfortunately, the right hand side still contains subdiagrams that are UV divergent for a scalar field theory with pointlike vertices, in $d=4$ dimensions. In particular, the third and last diagrams have tadpole-type subdiagrams.

\section{Incomplete Gamma function}
\label{app:GammaFunction}

For completeness, we collect here the properties of the (upper) incomplete Gamma function used in the main text. It is defined as\footnote{Note that the symbol $\Gamma$ is used for the Gamma function, the upper incomplete Gamma function, and the effective action, but the arguments allow to differentiate between them.}
\begin{equation}
  \Gamma(a,x) = \int_x^\infty dt \, t^{a-1} e^{-t}.
\end{equation}
This directly implies the derivative rule
\begin{equation}
  \frac{d}{dx} \Gamma(a,x) = - x^{a-1} e^{-x}.
\end{equation}
One can also show the following recursion relation,
\begin{equation}
  \Gamma(a+1,x) = a \Gamma(a,x) + x^{a} e^{-x}.
\end{equation}
For small $x$ one can use the expansion
\begin{equation}
  \Gamma(0,x) = - \gamma - \ln(x) - \sum_{n=1}^\infty \frac{(-x)^n}{n\times n!},
\end{equation}
and the recurrence relation
\begin{equation}
  \Gamma(-n,x) = \frac{1}{n!} \left( \frac{e^{-x}}{x^n} \sum_{k=0}^{n-1} (-1)^k (n-k-1)! x^k + (-1)^n \Gamma(0,x)\right).
\end{equation}
This implies for example
\begin{equation}
  \Gamma(-1,x) \approx \frac{e^{-x}}{x} + \gamma + \ln(x) - x \approx \frac{1}{x} -1 + \gamma + \ln(x),
\label{eq:IncompleteGammam1xAsympt}
\end{equation}
and
\begin{equation}
  x \Gamma(-1,x) \approx 1 + x \ln(x/e^{1-\gamma}).
\end{equation}
Here $\gamma \approx 0.57721$ is the Euler-Mascheroni constant.

The ordinary Gamma function is $\Gamma(a)=\Gamma(a,0)$. It has the well-known properties $\Gamma(x+1) = x \Gamma(x)$ for $x>0$ and $\Gamma(1)=1$. For integer $n>1$ this means $\Gamma(n) = (n-1)!$.

Around the origin one has the Laurent series
\begin{equation}
	\Gamma(x) = \frac{1}{x} - \gamma + \frac{1}{2} \left(\gamma^2+\frac{\pi^2}{6} \right) x + \ldots,
\end{equation}
while the expansion around $x=1$ is
\begin{equation}
	\Gamma(x) = 1 - \gamma(x-1) + \frac{1}{2}\left( \gamma^2 + \frac{\pi^2}{6} \right) (x-1)^2 + \ldots.
\end{equation}

\providecommand{\href}[2]{#2}\begingroup\raggedright\endgroup

\end{document}